\documentclass{iopjournal}

\usepackage{amsmath,amssymb,bm}
\usepackage[numbers,square,sort&compress]{natbib}
\usepackage{float}
\usepackage{etoolbox}

\definecolor{indigo}{HTML}{1A0040}
\colorlet{revisioncolor}{blue}

\usepackage{booktabs}
\usepackage{verbatim}
\usepackage{subcaption}
\usepackage{placeins}
\usepackage[normalem]{ulem}   
\usepackage{tikz}
\usepackage{lmodern}
\usetikzlibrary{shapes.geometric}
\newcommand{\openstar}[1][black]{%
  \tikz[baseline=-0.6ex]\node[star, star points=5, star point ratio=2.3,
        draw=#1, fill=white, line width=0.6pt, inner sep=1.4pt]{};}

\newcommand{\EP}{\mathrm{EP}}

\newcommand{\Pphi}{P_\varphi}

\begin{document}

\articletype{Paper}

\title{Global full-$f$ gyrokinetic analysis of energetic-particle-driven Alfv\'en eigenmodes and their phase-space drive with GYSELA}

\author{%
    S.~Raj$^{1}$, Y.~W.~Cho$^{1,2}$, Z.~S.~Qu$^{1,*}$,
    G.~Brochard$^{3}$, L.~Bramucci$^{4}$, P.~Donnel$^{5}$,
    R.~Bigu\'e$^{3,5}$, M.~Go$^{1}$, S.~Mazzi$^{5}$,
    G.~Dif-Pradalier$^{5}$, D.~Zarzoso$^{6}$, K.~Obrejan$^{5}$,
    V.~Grandgirard$^{5}$, Y.~Sarazin$^{5}$, K.~Lim$^{1}$~and~X.~Garbet$^{1,5}$%
}\par

\noindent\affil{$^{1}$School of Physical and Mathematical Sciences, Nanyang Technological University, Singapore}\par
\noindent\affil{$^{2}$Korea Institute of Fusion Energy (KFE), Daejeon, Republic of Korea}\par
\noindent\affil{$^{3}$Aix-Marseille Universit\'e, CNRS, PIIM UMR 7345 -- Laboratoire PIIM, site de Saint-J\'er\^ome, 52 avenue Escadrille Normandie-Niemen, 13397 Marseille cedex 20, France}\par
\noindent\affil{$^{4}$Consorzio RFX, Padova, Italy}\par
\noindent\affil{$^{5}$CEA, IRFM, F-13108 Saint Paul-lez-Durance, France}\par
\noindent\affil{$^{6}$Aix-Marseille University, CNRS, Centrale Med, M2P2, Marseille, France}\par
\noindent\affil{$^{*}$Author to whom any correspondence should be addressed.}\par
\email{zhisong.qu@ntu.edu.sg}\par

\noindent

\begin{abstract}
We investigate the linear stability of energetic-particle (EP) driven Alfv\'en eigenmodes and
the wave--particle interaction in phase space, using the newly implemented energetic-particle
module in the electromagnetic global full-$f$ gyrokinetic code GYSELA. The coupled
gyrokinetic--Maxwell system self-consistently evolves the EP species alongside thermal ions
and kinetic electrons, retaining finite-orbit-width and global profile effects. To mitigate the
relaxation of the initial distribution in global full-$f$ models, a canonical Maxwellian
initialisation is employed, which strongly reduces this relaxation and the associated change
of the drive. The framework is verified against the ITPA-TAE and the reversed-shear NLED-AUG
cases, including scans in EP temperature and density. The agreement with published
multi-code results is quantitative for the ITPA-TAE case and qualitative for the NLED-AUG
case, where a reversed-shear Alfv\'en eigenmode (RSAE) is found near $q_{\min}$. Beyond
verification, we use a constants-of-motion (CoM) diagnostic to project the gyrocentre
distribution and the local wave--particle power exchange ($P=-\mathbf J\cdot\mathbf E$) onto
invariant space. This identifies the orbit classes and resonances that interact with the mode. For the ITPA-TAE, deeply co-passing orbits dominate the drive. For the NLED-AUG RSAE,
counter-passing orbits drive the mode and co-passing orbits damp it, and the sign of the
exchange follows that of the linear EP drive.
These results provide a basis for future
global full-$f$ studies of nonlinear energetic-particle transport and mode saturation.
\end{abstract}

\section{Introduction}

Energetic particles (EP) produced by auxiliary heating systems and fusion reactions can resonantly interact with shear Alfv\'en waves, driving Alfv\'enic instabilities that significantly affect confinement and plasma performance \cite{Fu1989,ChenZonca2016}. Toroidal Alfv\'en eigenmodes (TAEs) arise from toroidal coupling between adjacent poloidal harmonics, which opens frequency gaps in the Alfv\'en continuum and permits discrete eigenmodes within otherwise heavily-damped frequency ranges.
In reversed-shear configurations, the safety factor exhibits a local minimum $q_{min}$, producing an extremum in the Alfv\'en continuum. This continuum structure allows discrete modes, known as reversed-shear Alfv\'en eigenmodes (RSAEs), to localise near $q_{min}$ and exhibit characteristic frequency evolution distinct from standard gap modes. Since these modes reside in spectral regions with reduced continuum damping, the stability of these eigenmodes is governed by the interplay between resonant fast-ion drive and residual background damping mechanisms. 
The accurate prediction of energetic-particle-driven Alfv\'enic activity therefore requires a self-consistent description of the thermal plasma that supports the waves as well as the wave-particle interaction in EP phase space. Several numerical approaches have been developed to study energetic-particle-driven Alfvénic instabilities. Hybrid MHD–kinetic models such as MEGA\cite{Todo1998PoP},  HMGC  \cite{Briguglio1995PoP}, HYMAGYC \cite{Vlad2021NLED}, NOVA-K \cite{Gorelenkov1999PoP}, CAS3D-K \cite{Konies2008AIP}, AE3D-K \cite{Spong2010CPP}, VENUS \cite{Cooper2011PPCF}, and XTOR-K~\cite{Brochard2020XTORK} treat energetic particles kinetically while retaining a fluid description for the thermal plasma. Gyrokinetic and particle-in-cell approaches, including LIGKA \cite{Lauber2007JCP}, GYGLES \cite{Mishchenko2009PoP}, ORB5 \cite{Jolliet2007CPC, Bottino2011PPCF, Lanti2020CPC}, EUTERPE~\cite{Kleiber2024CPC}, TRIMEG-GKX \cite{Lu2023PPCF}, gKPSP \cite{Cho2022PoP}, and GTC \cite{Wei2026GTC} have provided detailed insight into wave–particle interactions and finite-orbit-width effects. The gyrofluid code TAEFL \cite{Spong1992PFB} and FAR3D \cite{Varela2024FAR3d} offer an alternative reduced-model treatment of fast-ion closures.
These tools have successfully reproduced many linear properties of shear Alfvén modes and have played a central role in benchmark studies. 
Most of these results are based on a $\delta f$ approach, in which the ``equilibrium" state is removed, and a Krook operator is used to control the departure.

In contrast, a global full-$f$ gyrokinetic model evolves the complete distribution function of all species self-consistently. 
Electromagnetic field perturbations cause guiding-centre drifts that redistribute particles in phase space.
By retaining both the perturbation and the equilibrium state, the full-$f$ electromagnetic formulation allows nonlinear feedback between the equilibrium and the perturbation over long simulation times, which is difficult for $\delta f$ methods that are subject to numerical blow-up in that limit \cite{Chen2026canonical}. GYSELA~\cite{Grandgirard2016, Qu2026EMGYSELA} (for GYrokinetic SEmi-LAgrangian) provides such a full-$f$ electromagnetic framework by solving the coupled gyrokinetic Vlasov--Maxwell system in toroidal geometry without assuming scale separation. The full-$f$ framework is capable of capturing source-sink balance and the nonlinear redistribution of energetic particles in phase space at the same time.
However, the present work focuses on linear validation against multi-code benchmarks.

In toroidal geometry, a simple local Maxwellian that depends only on radius and energy
is not an exact kinetic equilibrium, as conservation of canonical toroidal momentum leads to redistribution through finite-orbit-width effects ~\cite{Brizard2011}. As a result, an initial distribution that is not a kinetic equilibrium can undergo spurious relaxation and flattening of invariant-space gradients before the linear phase of the instabilities, altering the effective resonant drive~\cite{Angelino2006, Idomura2003NF, DifPradalier2008PoP}. This effect is particularly pronounced in full-$f$ simulations using a local Maxwellian for energetic particles and has been shown to reduce the linear growth rate by approximately 40\% for $400$ keV energetic particles in the ITPA-TAE case~\cite{Lu2023PPCF}. To mitigate this effect, we initialise the plasma using a canonical Maxwellian consistent with the invariants of motion so as to reduce initial relaxation. 
\\
However, a small degree of profile adjustment remains unavoidable in a fully kinetic treatment. In a full-$f$ simulation the EP distribution itself is evolved, and for energetic particles the dominant effect is their finite orbit width. An initial distribution that is not exactly a function of the invariants of motion is therefore redistributed along the wide EP orbits. The canonical Maxwellian removes most of this redistribution. The small part that remains comes from the numerical initial distribution following the invariants only approximately, and it is phase-mixed along the particle trajectories within one or two EP transit times, which shows up as a small early relaxation of the EP profile~\cite{Lu2023PPCF}.
In addition, because the canonical Maxwellian is prescribed in terms of $\Pphi$, its real-space density is already at $t=0$ a finite-orbit-width smoothing of the prescribed profile, and this smoothing grows with the energy of the particles.
Since the TAE drive is set by the real-space EP gradient at the mode location, the EP fraction was tuned separately for each energy so that the real-space density gradient near $r/a \approx 0.5$ approximately matches that of the ITPA-TAE reference profile~\cite{Konies2018NF, Lu2023PPCF}.
\\
The purpose of the present work is to establish and validate a global full-$f$ kinetic
framework for investigating the wave--particle interaction in phase space of
energetic-particle-driven Alfv\'en eigenmodes, using the electromagnetic gyrokinetic code
GYSELA. While prior studies have successfully captured linear mode properties using $\delta f$
approaches, a full-$f$ formulation is well suited to capturing the self-consistent nonlinear
feedback between the equilibrium and the wave fields over long transport timescales. As a
first step, we verify the linear mode frequency, growth rate and spatial structure, and
identify the mode against the Alfv\'en continuum, in both the ITPA-TAE~\cite{Konies2018NF} and the
reversed-shear NLED-AUG~\cite{Vlad2021NLED} configurations. Furthermore, using the EPCoM
framework~\cite{BrochardEPCoM, Brochard2025NF}, we apply a constants-of-motion (CoM) phase-space diagnostic that
projects the gyrocentre distribution and the wave--particle power exchange
($P=-\mathbf J\cdot\mathbf E$) onto invariant space. This diagnostic identifies the orbit
classes responsible for drive and damping, and shows that the sign of the exchange is set by
the invariant-space gradients rather than by the radial density gradient alone.

The remainder of the paper is organised as follows: Section \ref{sec:em_implementation} provides the formulation and code overview, summarises the electromagnetic model and the mixed-variable implementation used in GYSELA, including mitigation of cancellation-related issues. Section~\ref{sec:refcases} presents the linear benchmark results and the constants-of-motion diagnostic, followed by the conclusions in Section \ref {sec:conclusions}.\\

\section{Electromagnetic formulation and Energetic Particle implementation in GYSELA}
\subsection{Overview of equations}
\label{sec:em_implementation}
\noindent
GYSELA is a global full-$f$ gyrokinetic code that evolves the complete gyrocentre distribution function of all plasma species in five-dimensional phase space $(\textbf{X},p_\parallel,\mu)$ in toroidal geometry where \textbf{X} is the gyrocentre position in the configurational space, $p_{\parallel}$ is the canonical velocity, and $\mu$ is the magnetic moment.
The spatial coordinates are $(r,\theta,\varphi)$, where $r$ is a radius-like flux label, $\theta$ is the generalised poloidal angle (does not need to be the geometrical angle or the straight-field-line angle), and $\varphi$ is the toroidal angle, same as the geometrical toroidal angle. 
In principle, an overhead bar should be added to all these variables to distinguish them from guiding centre variables.
Throughout the paper an overhead bar on a distribution function, potential or density denotes a gyrocentre 
quantity, while the phase-space coordinates themselves are written without a bar for simplicity.

In its electromagnetic extension, the code self-consistently solves the coupled gyrokinetic and Maxwell equations. The code simultaneously treats three plasma species: thermal ions, fully kinetic electrons, and an energetic-particle (EP) species that is evolved on an independent velocity-space grid.

\noindent
For each species $s$, the gyrocentre distribution function $\bar{F}_s$ evolves 
according to the conservative form of the full-$f$ electromagnetic gyrokinetic 
Vlasov equation,
\begin{equation}
\partial_t \left( B_{\parallel s}^* \bar{F}_s \right)
+ \nabla \cdot \left( B_{\parallel s}^* \dot{\mathbf{X}} \bar{F}_s \right)
+ \partial_{p_\parallel} \left( B_{\parallel s}^* \dot{p}_\parallel \bar{F}_s \right)
= B_{\parallel s}^* \mathcal{C}(\bar{F}_s),
\label{eq:gk_conservative}
\end{equation}
where $\mathcal{C}$ denotes the collision operator, Krook operator, and source 
terms, and the overhead dot represents the total time derivative along the 
gyrocentre trajectory. 

The effective 
magnetic field is given by $\mathbf{B}_s^* = \nabla \times \mathbf{A}_s^*$, where the 
modified vector potential is defined as
\begin{equation}
\mathbf{A}_s^* = \mathbf{A} 
+ \left( \frac{m_s \, p_\parallel}{q_s} + \mathcal{J}[A_{\parallel S}] \right) 
\mathbf{b},
\label{eq:Astar}
\end{equation}
with $\mathbf{A}$ the equilibrium magnetic vector potential 
($\mathbf{B} = \nabla \times \mathbf{A}$ the equilibrium magnetic field), 
$\mathbf{b} = \mathbf{B}/|\mathbf{B}|$ the unit vector along $\mathbf{B}$, 
$m_s$ and $q_s$ the mass and charge of species $s$, $p_\parallel$ the canonical 
parallel velocity, and $\mathcal{J}[\cdot]$ the gyro-average operator. 
GYSELA uses the mixed-variable scheme~\cite{Mishchenko2017} to mitigate the so-called cancellation problem.
Here, $A_{\parallel S}$ is the symplectic component of the parallel vector 
potential arising from the mixed-variable decomposition 
$A_\parallel = A_{\parallel S} + A_{\parallel H}$, where $A_{\parallel H}$ is 
the Hamiltonian component maintained small by the pullback transformation. 
Accordingly, $\mathcal{J}[A_{\parallel S}]$ is the symplectic parallel vector 
potential gyro-averaged at the gyrocentre position $\mathbf{X}$, encoding the 
finite-Larmor-radius contribution of $A_{\parallel S}$ to the modified magnetic 
geometry. The scalar Jacobian is obtained by projecting $\mathbf{B}_s^*$ onto 
$\mathbf{b}$:
\begin{equation}
B_{\parallel s}^* = B 
+ \left( \frac{m_s \, p_\parallel}{q_s} + \mathcal{J}[A_{\parallel S}] \right)
\frac{\mathbf{b} \cdot \nabla \times \mathbf{B}}{B},
\label{eq:Bstar}
\end{equation}
where $B = |\mathbf{B}|$.

The gyrocentre equations of motion are given by
\begin{equation}
B_{\parallel s}^* \dot{\mathbf{X}}
=
\frac{\mathbf{B}_s^*}{m_s} \partial_{p_\parallel} \bar H
-
\frac{\nabla \bar H \times \mathbf{B}}{q_s B},
\label{eq:motion_gyrocenter_position}
\end{equation}
\begin{equation}
m_s B_{\parallel s}^* \dot{p}_\parallel
=
- \mathbf{B}_s^* \cdot \nabla \bar H
- q_s B_{\parallel s}^* \partial_t A_{\parallel S},
\label{eq:motion_gyrocenter_velocity}
\end{equation}
where the Hamiltonian is written as
\begin{equation}
\bar H = \bar H_0 + \bar H_1,
\end{equation}
with
\begin{equation}
\bar H_0 = \frac{m_s}{2} p_\parallel^2 + \mu B,
\qquad
\bar H_1 = q_s \left( \bar \Phi - p_\parallel \bar A_{\parallel H} \right).
\end{equation}

The gyrocentre velocity-space integration measure is
\begin{equation}
dW = 2\pi B_{\parallel s}^* \, dp_\parallel \, d\mu,
\label{eq:volume_gyrocenter}
\end{equation}
which includes the phase-space Jacobian $B_{\parallel s}^*$ associated with the gyrocentre coordinate transformation.

The electrostatic potential $\Phi$ is obtained from the gyrokinetic quasi-neutrality condition
in its long-wave-length limit, given by
\begin{equation}
- \sum_s \nabla \cdot 
\left(
\frac{m_s n_{0,s}}{B^2} \nabla_\perp \Phi
\right)
=
\sum_s q_s \bar n_s ,
\label{eq:quasineutrality}
\end{equation}
where $n_{0,s}$ is the equilibrium density of species $s$, the left-hand side represents the polarisation charge from finite Larmor radius effects,
and the gyrocentre density on the right-hand side is
\begin{equation}
\bar n_s
=
2\pi\int d\mu \, dp_\parallel \,
\mathcal{J}^T
\left[
B_{\parallel s}^* \bar F_s
\right],
\end{equation}
with $\mathcal{J}^T$ denoting the adjoint gyroaveraging operator.

The parallel vector potential is obtained from the mixed-variable form of Amp\`ere's law,
\begin{equation}
- \nabla_\perp^2 A_{\parallel H}
+
\sum_s
\frac{\mu_0 q_s^2 n_{0,s}}{m_s}
A_{\parallel H}
=
\mu_0 \sum_s \delta \bar J_{\parallel s}
+
\nabla_\perp^2 A_{\parallel S},
\label{eq:ampere_mixed}
\end{equation}
where $\mu_0$ is the vacuum permeability and the perturbed parallel current is
\begin{equation}
\sum_s \delta \bar J_{\parallel s}
=
\sum_s
q_s
\int d\mu \, dp_\parallel \,
\mathcal{J}^T
\left[
p_\parallel B_{\parallel s}^* \bar F_s
\right]
-
J_{\parallel 0},
\end{equation}
with $J_{\parallel 0}$ the equilibrium parallel current density.

Both \eqref{eq:quasineutrality} and \eqref{eq:ampere_mixed} are solved by the Fourier-based finite difference solver detailed in Appendix \ref{app:FFT2D}.

The parallel canonical momentum is related to the canonical parallel velocity through
\begin{equation}
p_\parallel
=
v_\parallel
+
\frac{q_s}{m_s}
\left(
A_{\parallel H}
+
A_{\parallel S}
-
\bar{A}_{\parallel S}
\right),
\end{equation}
and the symplectic component evolves according to
\begin{equation}
\partial_t A_{\parallel S} = -\nabla_\parallel \Phi + \partial_t A_{\parallel N},
\label{eq:AparS_evolution}
\end{equation}
where the first term $-\nabla_\parallel \Phi$ corresponds to the ideal Ohm's law
(consistent with the description of  \cite{Qu2026EMGYSELA}), and
$A_{\parallel N}$ is a non-ideal correction term whose slope is updated each time step
as $\Delta A_{\parallel N}^{(i+1)} = \Delta A_{\parallel N}^{(i)} + A_{\parallel H}^{(i)}/\Delta t$,
where $\Delta t$ is the time step and the superscript $(i)$ labels the time index,
extrapolating the non-ideal contribution from the recent history of $A_{\parallel H}$.
This reduces $A_{\parallel H}$ from $O(\Delta t)$ to $O(\Delta t^2)$ compared with the
original mixed-variable scheme, significantly improving stability for nearly-electrostatic
modes such as drift waves and geodesic acoustic modes, without sacrificing accuracy for
Alfv\'enic perturbations.

The Vlasov equation is solved using a semi-Lagrangian scheme combined with Strang
splitting in phase space, allowing accurate treatment of parallel streaming and magnetic
drifts with controlled numerical diffusion.
Energetic particles are modelled in GYSELA as an additional kinetic species
with mass $m_{EP}$ and charge $q_{EP}$, evolved self-consistently with the bulk
plasma on the same five-dimensional phase space
$(\mathbf{X},p_\parallel,\mu)$.
Their gyrocentre distribution function $\bar{F}_{EP}$ obeys the same conservative
gyrokinetic Vlasov equation~\eqref{eq:gk_conservative} as the thermal
species, with the same gyrocentre equations of motion and modified magnetic
field $\mathbf{B}^*_s$.
\\
The EP population contributes to the field equations through its perturbed density and parallel current, allowing energy exchange between energetic particles and
plasma waves.
In particular, gradients of the EP distribution function in phase space
provide the free energy that can drive shear Alfv\'en instabilities such as
toroidal Alfv\'en eigenmodes (TAEs).
Because the EPs are evolved kinetically together with the bulk plasma,
wave--particle interaction arises naturally through the resonant condition
between the particle orbit frequencies and the mode frequency, enabling global
simulations of energetic-particle-driven instabilities and their nonlinear
interaction with background turbulence.

Each species carries its own velocity-space grid, defined independently 
in the parallel velocity $v_\parallel$ and magnetic moment $\mu$ 
coordinates. All grids are nevertheless expressed in a single set of reference units built from the main thermal ion species, which therefore serves as the reference against which the energetic-particle grid is extended. The $v_\parallel$ grid spans 
$[- N_{v,s}\,v_{\mathrm{th,s}},\, + 
N_{v,s}\,v_{\mathrm{th,s}}]$, where $v_{\mathrm{th,s}} = \sqrt{T_{\mathrm{ref}}/m_{s}}$ is 
the reference thermal velocity and $T_{\mathrm{ref}}$ the reference temperature, taken to be the main-ion temperature at mid-radius. The $\mu$ grid extends from $0$ to 
$L_{\mu,s}\,\mu_{ref}$, in units of $ \mu_{ref}= T_i/B_{\mathrm{ref}}$, where $B_{\mathrm{ref}}$ is the reference magnetic field taken as the on-axis field $B_0$, and $\mu_{ref}$ corresponds to a purely perpendicularly moving main ion at the thermal energy.

The dimensionless extents $N_{v,s}$ and $L_{\mu,s}$ are species-dependent and are specified for each benchmark case below. For the thermal species they need only cover a few thermal widths of the reference distribution. For the energetic particles, however, the same reference units correspond to a much narrower fraction of the fast-ion distribution, since
\begin{equation}
\frac{v_{\parallel,\mathrm{EP}}}{v_{\mathrm{th,ref}}}
\sim
\left(\frac{T_{\mathrm{EP}}}{T_i}\frac{m_i}{m_{\mathrm{EP}}}\right)^{1/2},
\qquad
\frac{\mu_{\mathrm{EP}}}{\mu_{ref}}
\sim
\frac{T_{\mathrm{EP}}}{T_i}.
\label{eq:ep_grid_scaling}
\end{equation}
The EP grid is therefore extended by these factors relative to the main-ion grid, so that the suprathermal population is resolved to a comparable number of its own thermal widths while all species remain on a common set of reference units.

It is to be noted that in the present work we have restricted the simulations to the 
collisionless limit, and collisions will be included in future work.

\subsection{Canonical Maxwellian initialisation and mitigation of radial profile relaxation}
\label{sec:canonical_maxwellian}
In global full-$f$ gyrokinetic simulations, a distribution that depends only on the local radial coordinate does not generally represent a true
kinetic equilibrium.
This is particularly important for energetic-particle (EP) populations,
whose guiding-centre orbits possess a finite radial width.

In toroidal geometry, the unperturbed guiding-centre motion conserves
the total energy $E$, the magnetic moment $\mu$, and the canonical
toroidal momentum
\begin{equation}
P_\varphi = m_{EP} \frac{I}{B} v_\parallel - q_{EP} \psi_p ,
\label{eq:Pphi}
\end{equation}
where $I(\psi) = RB_\varphi$ and $\psi_p$ is the poloidal magnetic flux.
Because of the conservation of $\Pphi$, particles initialised at a given flux
surface can travel across neighbouring surfaces during their guiding-centre
motion.
As a result, an initial distribution depending only on the local flux
coordinate $r$, and the total EP energy $E = \frac{1}{2}m_{EP} v_{\parallel}^2 + \mu B$, 
\begin{equation} 
F_M= n_{\mathrm{EP}}(r) \left(\frac{m_{EP}}{2\pi T_{\mathrm{EP}}(r)}\right)^{3/2} \exp\left(-\frac{E}{T_{\mathrm{EP}}(r)}\right), 
\label{eq:FM} 
\end{equation} 
is not constant along particle trajectories, since $r$ is not an invariant of the gyrocentre motion.
This leads to an initial  redistribution phase in which the
EP density profile relaxes before the onset of the linear instability.
Such relaxation modifies the effective invariant-space gradients
responsible for the energetic-particle drive.

To reduce this effect, GYSELA employs a canonical Maxwellian
initialisation, where the radial dependence is expressed in terms of a
drift-orbit–averaged flux coordinate $\bar r$ rather than the local
guiding-centre position.
The canonical flux coordinate is obtained by correcting the local
poloidal flux $\psi$ for the finite-orbit-width contribution associated
with parallel motion.

Following the implementation in GYSELA, originally introduced in ORB5~\cite{Angelino2006}, the orbit-corrected poloidal flux is defined as

\begin{equation}
\psi_{\mathrm{bar}}
=
\psi-\frac{\sqrt{\hat A_s}}{\hat Z_s}
\left(
\frac{I}{B}v_\parallel
-
\frac{I_0}{B_0}\bar v
\right),
\end{equation}

where $\hat A_s$ and $\hat Z_s$ are the normalised mass and charge of
species $s$,  and
$I_0$ and $B_0$ are those quantities on axis.
The term proportional to $(B_\varphi/B)\,v_\parallel$ represents the canonical
momentum contribution associated with the particle's parallel motion.
The reference velocity $\bar v$ corresponds to the passing-particle
orbit velocity and is defined as
\begin{equation}
\bar v
=
\mathrm{sgn}(v_\parallel)
\sqrt{\frac{2(E-\mu B_0)}{m_s}}
\, \mathcal{H}(E-\mu B_{\max}),
\end{equation}
where $B_\text{max}$ is the maximum field strength on that flux surface and $\mathcal{H}$ is the Heaviside function.
This definition ensures that the correction is applied only to passing
particles, while trapped particles (for which $E < \mu B_{\max}$)
retain $\bar v = 0$.

The corrected flux $\psi_{\mathrm{bar}}$ is then mapped to the
corresponding radial coordinate
\begin{equation}
\bar r = r(\psi_{\mathrm{bar}}),
\end{equation}
using the equilibrium mapping between poloidal flux and radius.

Phase-space grid points whose orbit-corrected flux $\psi({\bar{r}})$ falls outside the domain $[0, \psi_{max}]$, where $\psi_{max}$ is the poloidal flux at the last closed flux surface, are remapped to the nearest boundary value rather than being removed.

The canonical Maxwellian is therefore written as
\begin{equation}
F_{\mathrm{CM}}(\bar r,E)
=
n_{\EP}(\bar r)
\left(\frac{m_{EP}}{2\pi T_{\EP}(\bar{r})}\right)^{3/2}
\exp\!\left(-\frac{E}{T_{\EP}(\bar{r})}\right).
\end{equation}

Because $\bar {r} $ remains approximately constant along unperturbed
guiding-centre trajectories, the distribution function is nearly
invariant during the initial phase of the simulation as long as $\bar H_1$ is small compared to $\bar H_0$.
This significantly reduces the artificial relaxation of the energetic
particle density profile and preserves the intended invariant-space
gradients responsible for the TAE drive.
Consequently, the linear growth rates measured in the ITPA benchmark
reflect the physical wave–particle interaction rather than a transient
adjustment of the initial distribution. In future work, EP distributions will be loaded directly in CoM coordinates from EPCoM \cite{BrochardEPCoM}.

\section{Benchmark cases and numerical setup}
\label{sec:refcases}
Two benchmark configurations are considered.
The first is the ITPA-TAE reference case, widely used for code-to-code comparisons in simplified tokamak geometry \cite{Konies2018NF}.
The second is the NLED-AUG on-axis case~\cite{Vlad2021NLED}, based on the
ASDEX Upgrade discharge \#31213 at $t = 0.84$~s, a strongly shaped,
reversed-shear equilibrium with experimental density and temperature
profiles and a peaked on-axis population of NBI deuterons. Together, the
two cases cover a simplified circular geometry and a shaped experimentally based configuration. All results presented here are taken in the linear phase of the modes. 

\subsection{Reference cases: ITPA-TAE}
The ITPA-TAE case is used as a controlled reference for energetic-particle (EP) driven TAEs, with extensive multi-code comparisons available in the literature.

\begin{figure}[!htbp]
    \centering
    \begin{subfigure}[b]{0.49\linewidth}
        \centering
        \includegraphics[width=\linewidth]{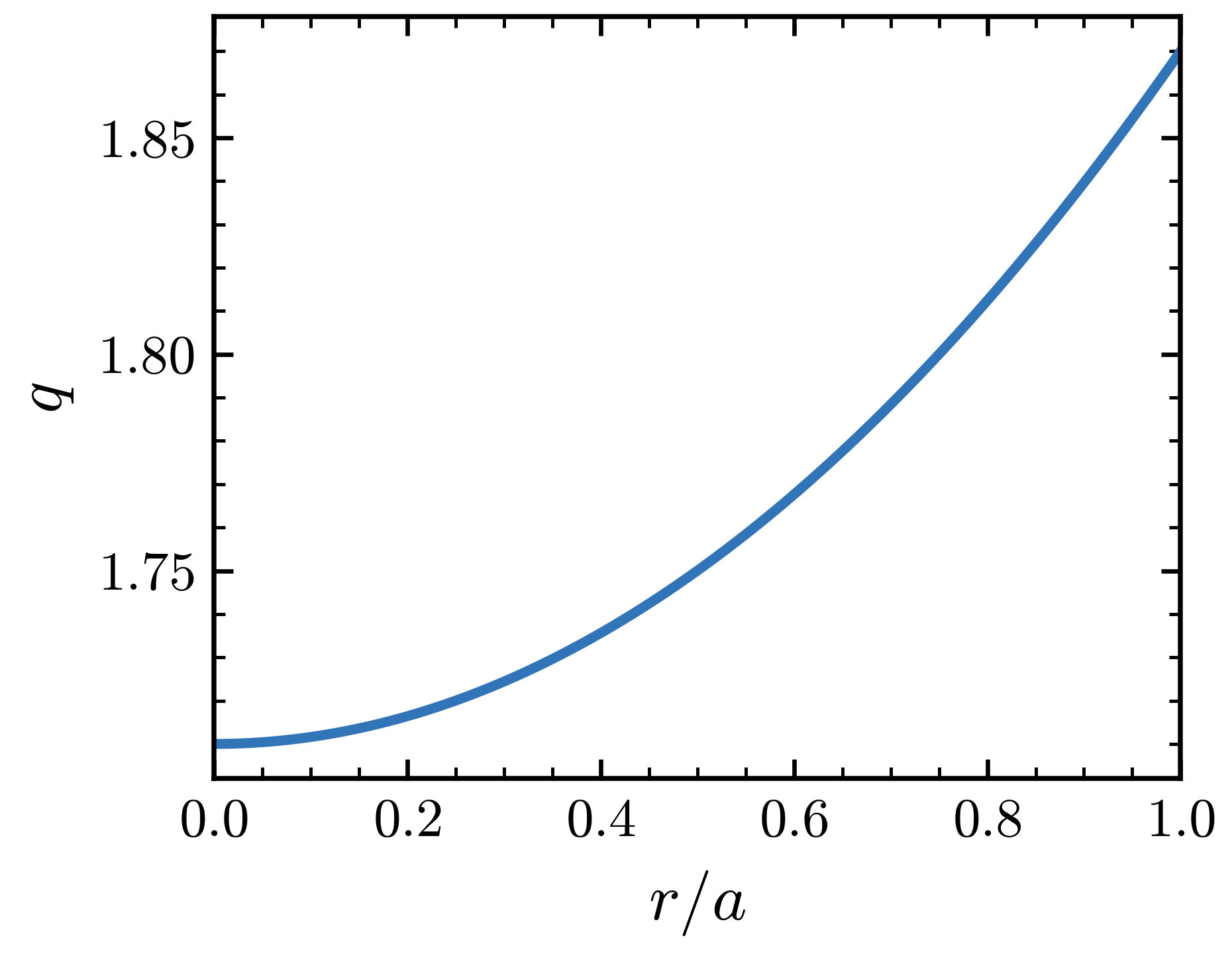}
        \caption{}
        \label{fig:itpa_safety_factor}
    \end{subfigure}
    \hfill
    \begin{subfigure}[b]{0.49\linewidth}
        \centering
        \includegraphics[width=\linewidth]{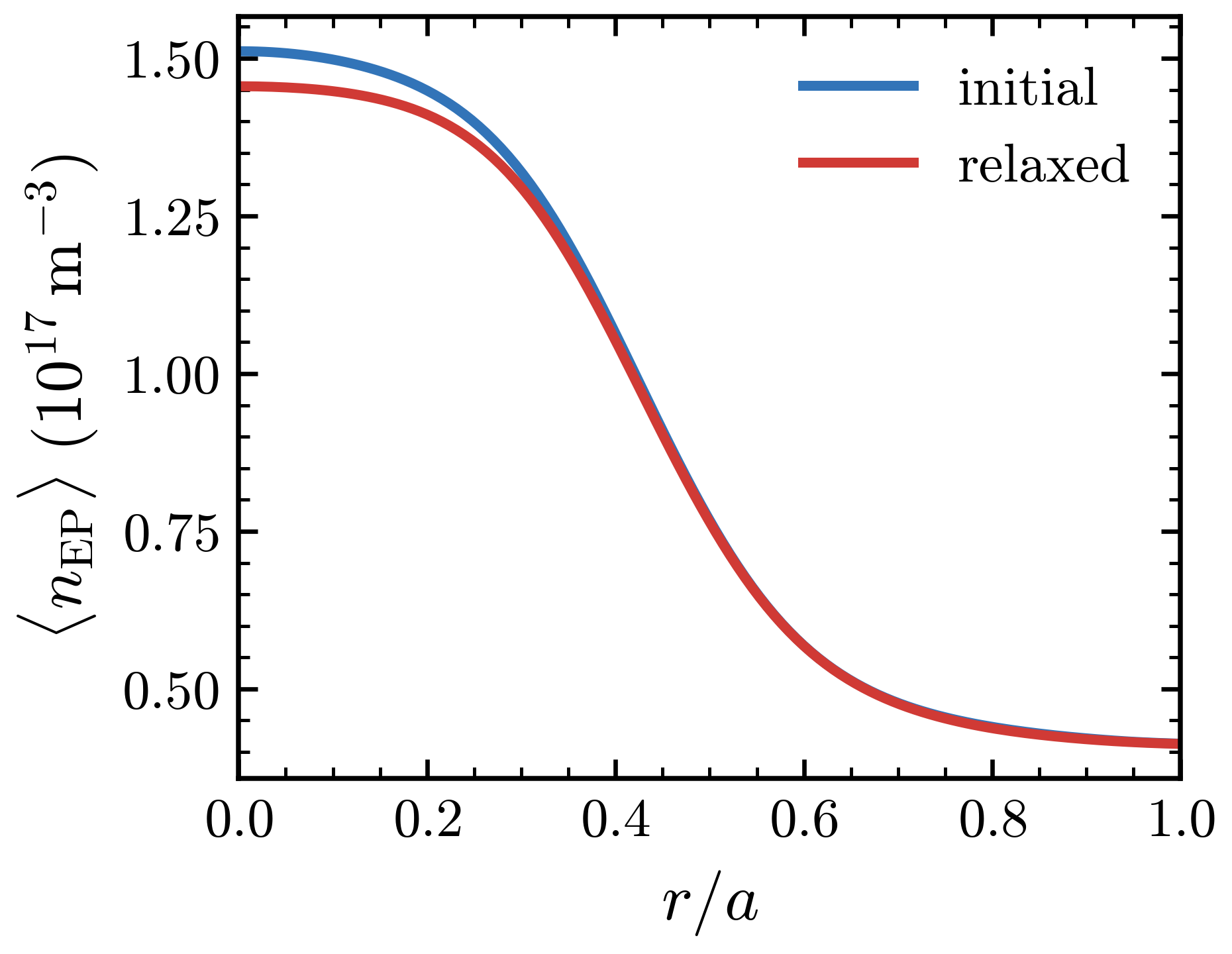}
        \caption{}
        \label{fig:itpa_density}
    \end{subfigure}
    \caption{ITPA-TAE profiles used in GYSELA: (a) safety factor
    (b) flux-surface-averaged energetic-particle density at the start of the
    simulation (blue) and after the initial relaxation (red),
    $T_{\rm EP} = 400$~keV.}
    \label{fig:itpa_profiles}
\end{figure}

For the ITPA-TAE configuration, we consider a tokamak geometry with concentric circular flux surfaces, consistent with common benchmark assumptions.
The safety factor profile as shown in Fig.~\ref{fig:itpa_safety_factor} is prescribed in \cite{Konies2018NF}
\begin{equation}
q(r) = 1.71 + 0.16 \left(\frac{r}{a}\right)^2 ,
\label{eq:itpa_q}
\end{equation}
where $a$ is the minor radius.

The energetic-particle density follows a smooth tanh-based form commonly used in ITPA-TAE benchmarks and recent electromagnetic gyrokinetic studies \cite{Lu2023PPCF},
\begin{equation}
n_{\EP}(r) = n_{\EP,0}\,c_3 \exp\!\left[-\frac{c_2}{c_1}\tanh\!\left(\frac{r-c_0}{c_2}\right)\right],
\end{equation}
where $n_{\EP,0}$ is the reference EP density and $c_0$--$c_3$ are shape coefficients controlling
the position, width and amplitude of the gradient region, and the energetic particle temperature, $T_{EP}$, is scanned over a wide range while bulk profiles and the equilibrium are kept fixed, and the equilibrium does not include the EP pressure, as in the benchmark definition \cite{Konies2018NF}. $T_{EP}$ is radially uniform.  The original benchmark uses $(c_0, c_1, c_2, c_3) = (0.49123, 0.298228, 0.198739, 0.521298)$ with a local Maxwellian\cite{Konies2018NF}. In the present work, we choose $\big({c_{0},c_{1},c_{2},c_{3}}\big)_{can}=\big(0.466 23, 0.170 42, 0.113 57, 0.521 298\big)$, where the subscript ``can'' indicates that these values are those used for the canonical Maxwellian initialisation of Ref.~\cite{Lu2023PPCF}. They are different from those of the original benchmark~\cite{Konies2018NF}, but would give the same radial density profile as the original benchmark after initialisation.

In this benchmark, the equilibrium and the bulk plasma profiles are fixed, while the EP
parameters are scanned. The background-ion density is kept flat at
$n_0 = 2.0\times10^{19}$~m$^{-3}$ (up to the fast-ion fraction), and quasi-neutrality is
maintained through the electron density, $n_e(r) = n_i + n_{\mathrm{EP}}(r)$. The electron
and ion temperatures are flat, $T_e = T_i = 1$~keV, and the EP density follows Eq.~(23) with
$n_{\mathrm{EP},0} = 1.44131\times10^{17}$~m$^{-3}$, the fast-ion fraction being adjusted
with $T_{\mathrm{EP}}$ as discussed below. This makes the case ideal to validate (i) the EP
drive implementation, (ii) the electromagnetic coupling through $A_\parallel$, and (iii) the
sensitivity of growth rates to the EP phase-space gradients.

\paragraph{Numerical parameters.}
The ITPA-TAE simulations are performed on a five-dimensional grid with 
$(N_r, N_\theta, N_\varphi, N_{v_\parallel}, N_\mu) = (127, 128, 16, 256, 63)$, 
giving a minor radius of $a = 928.59\,\rho_i$ and an aspect ratio $R/a = 10$, where
$\rho_i = m_i v_{\mathrm{th},i}/(q_i B_0)$ is the thermal ion Larmor radius.
The toroidal domain spans $\Delta\varphi = \pi/3$. A single $n=1$ harmonic would correspond to $n=6$ in the full torus.

Three species are evolved simultaneously: background hydrogen ions
($A_i = 1$, $Z_i = 1$), kinetic electrons ($Z_e = -1$) and energetic
deuterons ($A_{\rm EP} = 2$, $Z_{\rm EP} = 1$), where $A_s = m_s/m_p$ and
$Z_s = q_s/e$. The electrons use a reduced mass ratio $m_i/m_e = 200$
($A_e = 0.005$) instead of the physical value, which relaxes the time-step
limit set by the electron parallel streaming, $\Delta t\, k_{\parallel,\max} v_{\mathrm{th},e} \lesssim 1$ with $v_{\mathrm{th},e} = (T_e/m_e)^{1/2}$, so that $\Delta t_{\max} \propto m_e^{1/2}$ is
about $3$ times larger than with the physical electron mass.

Each species $s$ has its own velocity-space grid in $(v_\parallel, \mu)$.
The grids are expressed in common units based on the reference
temperature $T_{\rm ref}$, taken as the main-ion temperature at mid-radius,
and the on-axis magnetic field $B_0$. The parallel velocity $v_\parallel$ is measured in units of
$v_{{\rm th},s} = (T_{\rm ref}/m_s)^{1/2}$ and $\mu$ in units of
$\mu_{\rm ref} = T_{\rm ref}/B_0$. The $v_\parallel$ grid spans
$[-N_{v,s}\,v_{{\rm th},s},\, N_{v,s}\,v_{{\rm th},s}]$ and the $\mu$ grid
$[0,\, L_{\mu,s}\,\mu_{\rm ref}]$.
For the energetic particles $T_{\rm EP} \gg T_{\rm ref}$, and the same width requires
\begin{equation}
  N_{v,{\rm EP}} = N_v \left(\frac{T_{\rm EP}}{T_{\rm ref}}\right)^{1/2},
  \qquad
  L_{\mu,{\rm EP}} = L_\mu\,\frac{T_{\rm EP}}{T_{\rm ref}},
  \label{eq:ep_grid}
\end{equation}
where $N_v$ and $L_\mu$ are the extents used for the thermal species.
The thermal-species grids span $\pm 4\,v_{{\rm th},s}$ in $v_\parallel$ and
$0$--$8\,\mu_{\rm ref}$ in $\mu$. The EP grid follows
Eq.~\eqref{eq:ep_grid} at each temperature of the scan, so that it always
covers $\pm 4\,v_{\rm th,EP}$ and $0$--$8\,\mu_{\rm th,EP}$, with
$v_{\rm th,EP} = (T_{\rm EP}/m_{\rm EP})^{1/2}$ and
$\mu_{\rm th,EP} = T_{\rm EP}/B_0$. In particular at $T_{\rm EP} = 400$~keV this gives
$N_{v,{\rm EP}} = 80$ and $L_{\mu,{\rm EP}} = 3200$.

As shown in Fig.\ref{fig:itpa_density}, the initial profiles (blue curve) correspond to the canonical Maxwellian loaded at $t=0$ used to initialise the simulation, while the red curves represent the relaxed profiles. 
As a consequence, the effective gradients in the drive region are slightly modified compared with the initial profiles.

Since the energetic-particle drive depends sensitively on the gradients
of the distribution function in phase space, the effective drive after
relaxation may differ from that associated with the reference prescribed
fast-ion fraction. For this reason, some adjustment of the initial fast-ion fraction is introduced in the temperature scan in order to recover a comparable post-relaxation gradient level to that used in the benchmark specification. 
Increasing the fast-ion density fraction $n_f$ restores $\nabla n(r)$, since the density
is the velocity-space integral of $f$, scaling $f$ uniformly by $n_f$ scales the density
profile, and hence its gradient, by the same factor at every radius, independently of
the orbit width. This adjustment should therefore be interpreted as a compensation for the full-$f$ relaxation rather than an arbitrary tuning of the instability growth rate.

\begin{figure}[!htbp]
    \centering
    \includegraphics[width=0.7\linewidth]{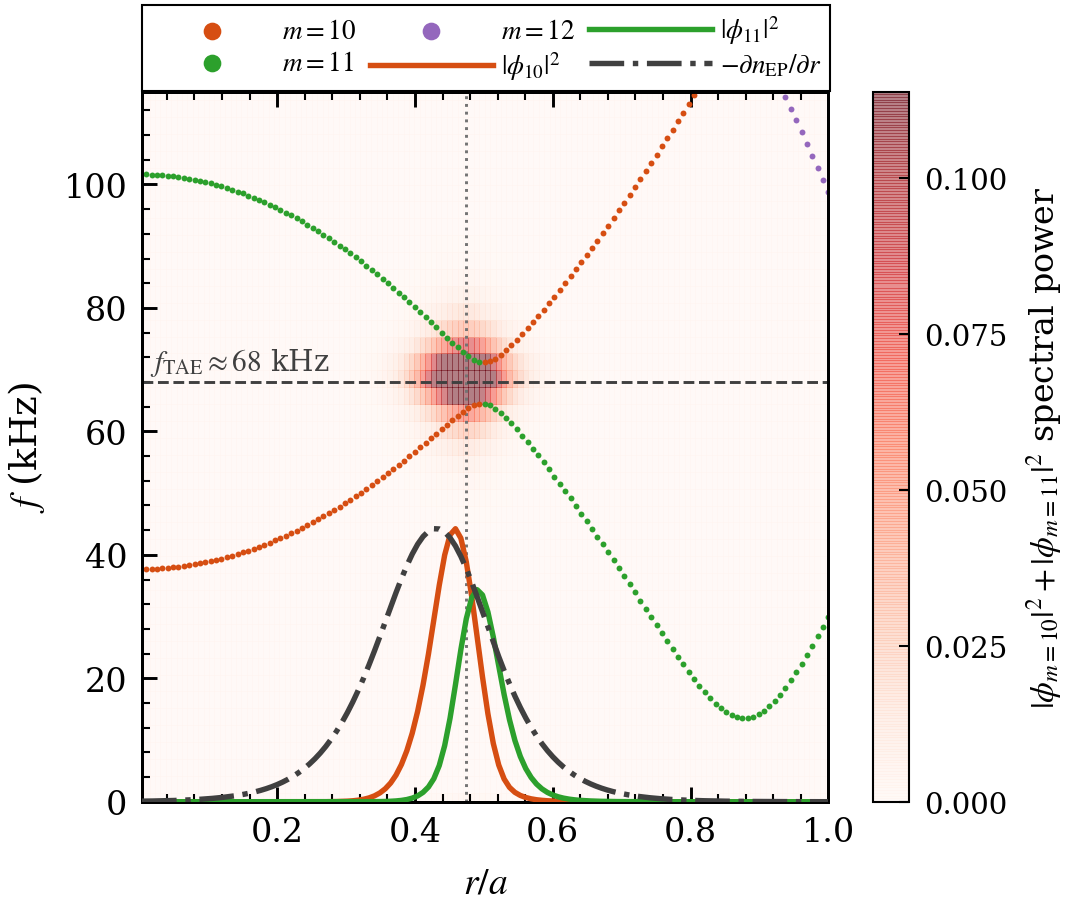}

\caption{ITPA-TAE case at $T_{\mathrm{EP}} = 400\,\mathrm{keV}$: shear Alfv\'en
continuum for $|n| = 6$ from ALCON~\cite{Deng2012NF} (dots), spectral power
$|\hat\phi_{m=10}|^2 + |\hat\phi_{m=11}|^2$ (colour), radial envelopes of the
two harmonics (solid) and $-\partial n_{\mathrm{EP}}/\partial r$ (dash-dotted,
arbitrary units). Harmonics are computed in the straight-field-line angle
$\theta^\star$. The dashed and dotted lines mark the mode frequency and the
radius of peak power.}
    \label{fig:itpa_continuum_harmonics}
\end{figure}

The profile shown in Fig. \ref{fig:itpa_density} is intentionally peaked and has a strong radial gradient in the drive region,
so it contains free energy that can be transferred to the wave through wave--particle resonance.
However, what matters for the drive is not only the real-space gradient $dn_{\mathrm{EP}}/dr$,
but the corresponding gradients in the constants-of-motion (CoM) space,
because resonances are organised by invariants (e.g., $P_\varphi$, $\lambda$, and $E$).
Therefore, Fig.~\ref{fig:itpa_density} should be interpreted as the real-space source of drive,
while the phase-space diagnostic presented in Fig.~\ref{fig:com_fluct_reso} shows which orbit classes actually access that free energy.

Fig.~\ref{fig:itpa_continuum_harmonics} shows the mode at $T_{\rm EP} = 400$~keV
together with the shear Alfv\'en continuum for $n = -6$ (the negative sign follows the toroidal-angle convention of GYSELA), computed with
ALCON~\cite{Deng2012NF} for $m = 10$, 11 and 12 in straight-field-line
coordinates. The $m = 10$ and $m = 11$ branches cross near $r/a \approx 0.5$,
where toroidal coupling opens a gap around $f \approx 68$~kHz. The spectral
power of the $m = 10$ and $m = 11$ harmonics in GYSELA is concentrated inside
this gap, so the mode is not absorbed by the continuum. A standard signature
of a TAE is a dominant coupled pair of neighbouring harmonics $(m, m+1)$,
together with smaller sidebands~\cite{Sadr2022}. The radial
envelopes show the pair $(10, 11)$ localised at the gap. The observed
harmonic structure therefore confirms that the instability is a
toroidicity-induced Alfv\'en eigenmode.

The run at $T_{\mathrm{EP}} = 400\,\mathrm{keV}$ provides the reference values for the eigenmode structure shown in Fig.~\ref{fig:itpa_continuum_harmonics} and the CoM diagnostics shown in Fig.~\ref{fig:com_fluct_reso}.

\begin{figure}[!htbp]
    \centering
    \begin{subfigure}[b]{0.49\linewidth}
        \centering
        \includegraphics[width=\linewidth]{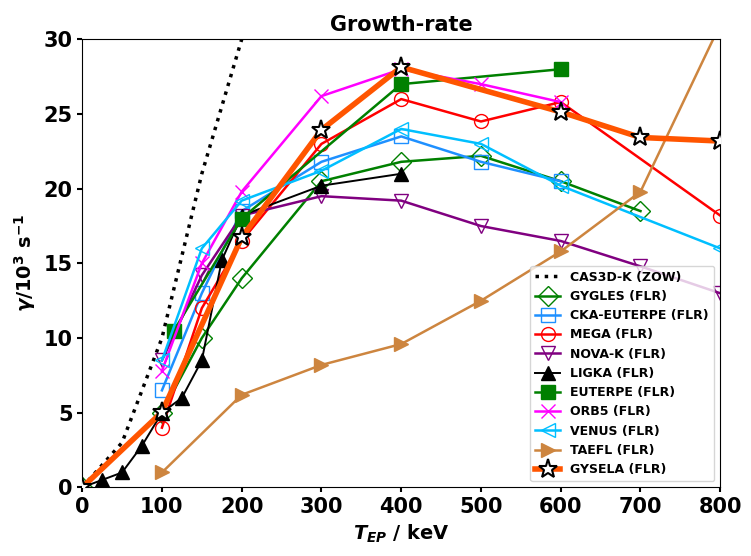}
        \caption{}
        \label{fig:itpa_gamma}
    \end{subfigure}
    \hfill
    \begin{subfigure}[b]{0.49\linewidth}
        \centering
        \includegraphics[width=\linewidth]{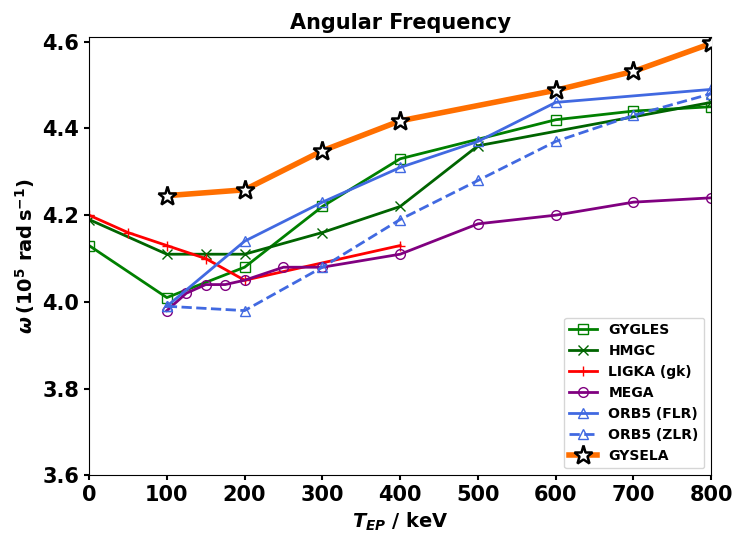}
        \caption{}
        \label{fig:itpa_omega}
    \end{subfigure}
    \caption{ITPA-TAE benchmark:
    (a) growth rate $\gamma$ versus EP temperature $T_{\mathrm{EP}}$
    compared to published benchmark results.
    (b) Real frequency $\omega$ versus fast-ion temperature
    $T_{\mathrm{EP}}$ compared to published benchmark results.
    The \protect\openstar[black] marks the GYSELA runs.}
    \label{fig:itpa_gamma_omega}
\end{figure}

\begin{figure}[!htbp]
\centering
\includegraphics[width=0.6\linewidth]{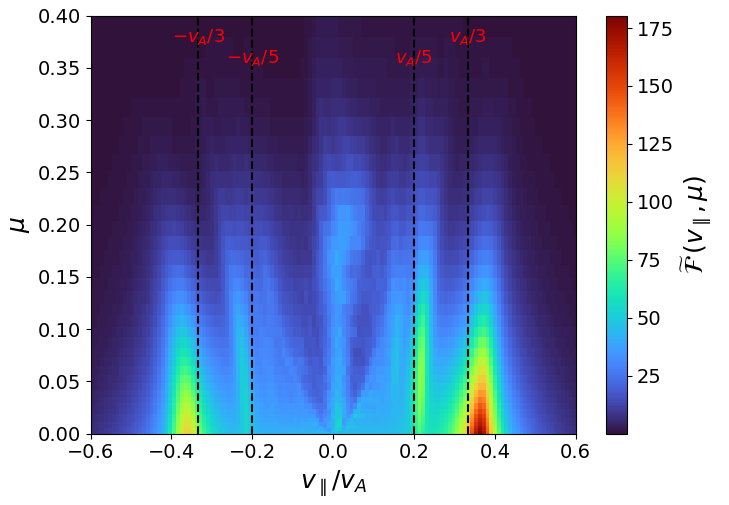}
\caption{ITPA-TAE benchmark: velocity-space resonance signatures at $T_{EP} = 400\,\mathrm{keV}$, shown as velocity-space RMS diagnostics (format aligned with published benchmark practice). Here $v_A = B_0/\sqrt{\mu_0 n_i m_i}$ is the on-axis Alfv\'en velocity.}
\label{fig:itpa_vspace}
\end{figure}

Fig.\ref{fig:itpa_gamma} compares the linear growth rate $\gamma$ as a function of EP temperature $T_{EP}$ against published benchmark trends.
The main result is that GYSELA reproduces the expected temperature dependence and remains within the multi-code spread.
Physically, increasing $T_{EP}$ modifies the EP distribution and orbit width,
which changes the resonant interaction between the particles and the TAE that can affect the drive.
As $\gamma$ depends on a delicate balance between EP drive and damping mechanisms, agreement with the benchmark trend in Fig.\ref{fig:itpa_gamma}
verifies the new EP module in the electromagnetic version of GYSELA \cite{Qu2026EMGYSELA}.

The higher growth rates of GYSELA compared with most codes are mainly attributed to the EP density normalisation. For each energy, the EP density was increased so that $\partial n_{EP}/\partial r$ at r/a = 0.5 matches the benchmark target after the initial relaxation.

The real frequency $\omega$ as a function of $T_{EP}$ is shown in Fig.\ref{fig:itpa_omega}. At 400 keV the GYSELA growth rate coincides with ORB5, and the frequency lies in the same TAE gap and is a few percent higher than the other codes over the whole scan, consistent with the increased EP density, since the TAE frequency increases with the EP fraction \cite{Konies2018NF}.

In contrast to the growth rate, the frequency is primarily determined by Alfv\'en gap physics and equilibrium properties.
The weak variation of $\omega$ across the scan confirms that the mode remains on the TAE branch
and that the electromagnetic field solver captures the correct shear-Alfv\'en response. The mode remains dominated by Alfvénic gap physics across the $T_{EP}$ scan and does not turn into an energetic-particle mode (EPM), the mode structure is therefore not expected to change significantly between the cases. We therefore take $T_{EP} = 400$ keV as a representative point and examine the poloidal harmonic structure in detail. Agreement of both the harmonic content and radial localisation with benchmark expectations
demonstrates that the global EM–EP capability captures the correct eigenmode structure.
\\
Velocity-space resonance signatures at $T_{\rm EP} = 400$~keV are shown in
Fig.\ref{fig:itpa_vspace} as the RMS of the toroidal fluctuation of the EP
distribution, $\tilde{F}(r, v_\parallel, \mu)$ of Eq.~\eqref{eq:vrms},
evaluated at the mid-radius where the mode amplitude peaks,
consistent with published benchmark studies~\cite{Konies2018NF}.
The colour encodes the configuration-space RMS of the
toroidal fluctuation of the EP distribution function,
\begin{equation}
\mathcal{\tilde{F}}(r, v_\parallel,\mu)
=
\left[
\frac{1}{N_\theta N_\varphi}
\sum_{\theta,\varphi}
\delta F_{EP}^2(\mathbf{X},v_\parallel,\mu)
\right]^{1/2},
\label{eq:vrms}
\end{equation}
where $\delta F_{EP} = F_{EP} - \langle F_{EP}\rangle_\varphi$
is the departure from the toroidal mean and  $N_\theta$ and $N_\varphi$
are the numbers of grid points in $\theta$ and $\varphi$. It measures the non-axisymmetric part of the distribution function, which encodes the wave--particle interaction. As GYSELA evolves the
complete distribution and has no separately prescribed equilibrium, the
axisymmetric part of $F_{\rm EP}$ itself evolves, in particular during the
initial relaxation shown in Fig.\ref{fig:itpa_profiles}(b). The toroidal
average $\langle F_{\rm EP}\rangle_\varphi$ is exactly the $n = 0$ component of
$F_{\rm EP}$, which is the instantaneous, relaxed equilibrium, and
$\delta F_{\rm EP} = F_{\rm EP} - \langle F_{\rm EP}\rangle_\varphi$ is the sum of
the $n \neq 0$ components.

The dashed vertical lines at \(v_\parallel/v_A=\pm 1/3\) and \(\pm 1/5\) indicate higher-order wave–particle resonances of passing fast ions with the TAE, corresponding to transit harmonics \(\ell=1\) and \(\ell=2\), respectively~\cite{Konies2018NF, Betti1992, Heidbrink2008}. Here, \(\ell\) labels the harmonics of the fast-ion transit motion, rather than the poloidal mode number \(m\) of the TAE. Each transit harmonic contributes an additional \(\ell v_\parallel/(qR)\) to the fundamental transit frequency $(m/q-n)v_\parallel/R$, yielding resonant velocities \(|v_{\parallel,\mathrm{res}}|\simeq v_A/(2\ell+1)\), with \(\ell=0\) corresponding to the fundamental resonance.

This plot confirms that the instability is driven by a resonant EP
population rather than by numerical artefacts. In physical terms,
Fig.\ref{fig:itpa_vspace} identifies the velocity regions where
wave--particle energy exchange is strongest. This provides a bridge 
to the CoM diagnostic introduced in the next section, which 
reorganises these resonances in invariant space to identify the 
responsible orbit classes and specific resonances.

\subsection{Constants-of-motion phase-space diagnostic}
A central challenge in EP-driven Alfv\'enic instabilities is that the resonance condition is written in terms of orbit frequencies, which depend on invariants such as $(P_\varphi,\lambda, E)$ \cite{Brochard2025NF, Brochard2024, BrochardEPCoM}. Here $P_\varphi$ is the canonical toroidal momentum of Eq.~(\ref{eq:Pphi}), conserved by axisymmetry and encoding the radial extent of the guiding-centre orbit. $\lambda=\mu B_0/E$ is the pitch angle parameter. $E$ is the equilibrium kinetic energy of the gyrocentre, conserved along the unperturbed guiding-centre trajectories.
Therefore, even if the real-space profile (e.g., EP density/pressure vs radius) looks strongly peaked, only a subset of particle orbits may actually resonate with the mode and contribute to the drive.

The CoM diagnostic directly addresses this issue by identifying which orbit classes satisfy the TAE resonance condition and whether those orbits intersect a steep or gentle gradient of $F$. The drive is governed by $\big[n\partial F/\partial \Pphi + \omega\big(\partial F/\partial E - (\lambda/E)\partial F/\partial \lambda\big)\big]$ evaluated at the resonance that governs the wave--particle energy exchange \cite{BerkBreizman1990, Pinches1998, Zonca2015, Brochard2025NF, BrochardEPCoM}

The linear wave--particle resonance condition may be written as
\begin{equation}
    \omega_0
    =
    n\bar{\omega}_d
    +
    p\,\omega_b ,
    \label{eq:res_trapped}
\end{equation}
where $\omega_0=2\pi f$ is the mode angular frequency,
$\bar{\omega}_d$ is the orbit-averaged toroidal
frequency, $\omega_b$ denotes the bounce frequency for trapped
particles and the corresponding transit frequency for circulating
particles, $p$ is the bounce/transit harmonic index,
and $n$ is the toroidal mode number~\cite{ChenZonca2016}.

\begin{figure}[!htbp]
    \centering
    \begin{subfigure}[b]{0.49\textwidth}
        \centering
        \includegraphics[width=\linewidth]{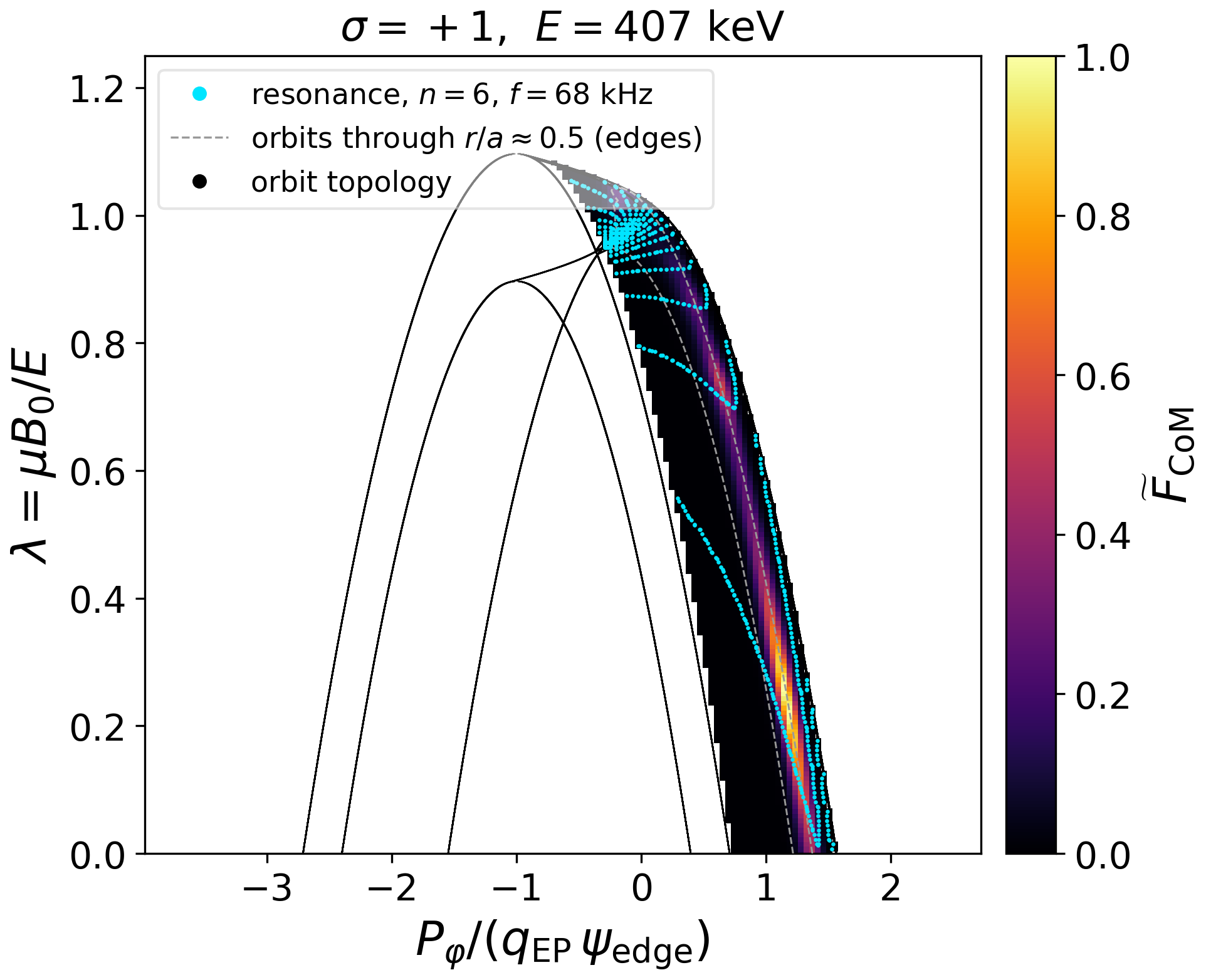}
        \caption{}
        \label{fig:com_fluct_co}
    \end{subfigure}
    \hfill
    \begin{subfigure}[b]{0.49\textwidth}
        \centering
        \includegraphics[width=\linewidth]{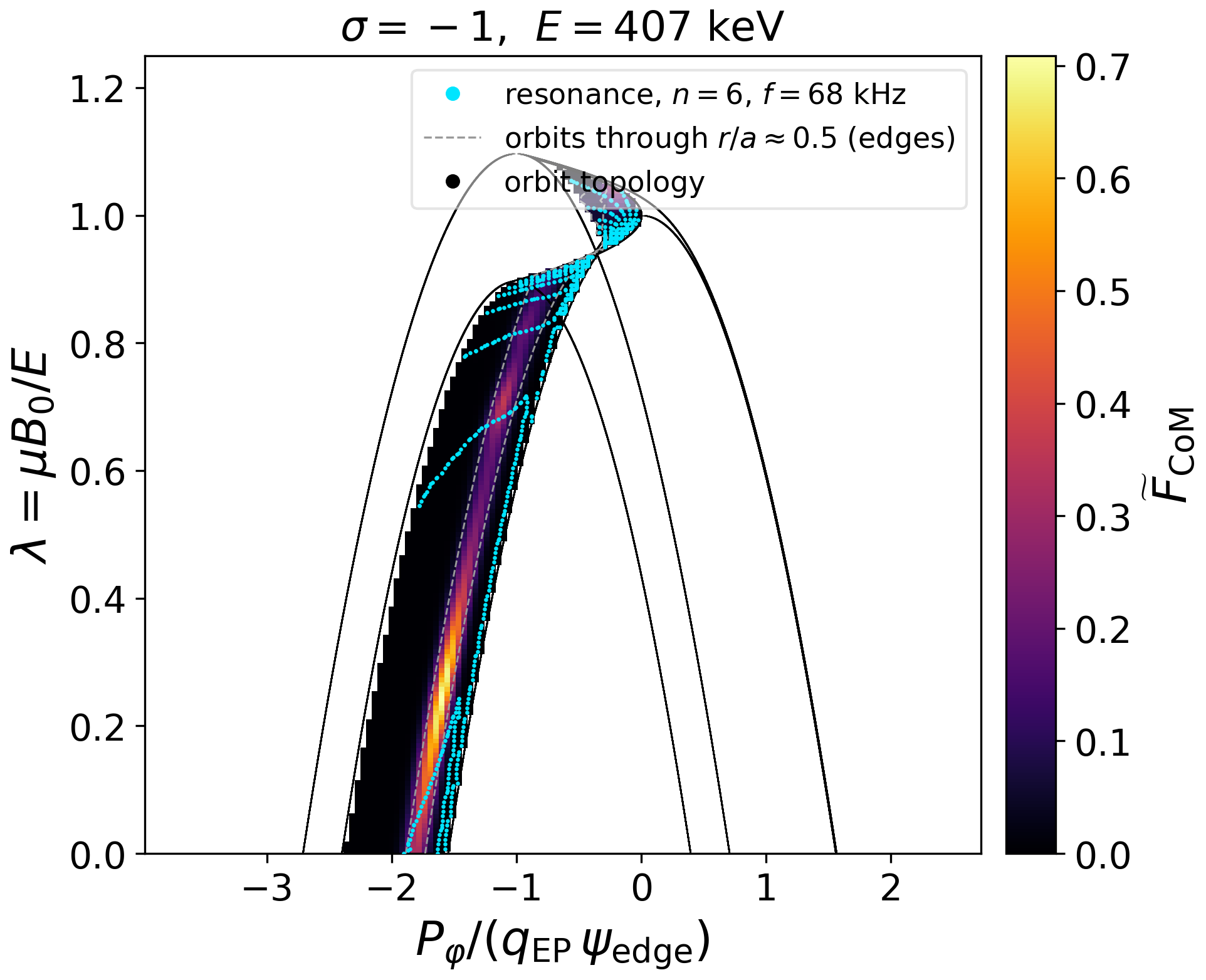}
        \caption{}
        \label{fig:com_fluct_counter}
    \end{subfigure}
    \caption{Fluctuation maps $\tilde{F}_{CoM}$ in constants-of-motion space at fixed energy $E=407\,\mathrm{keV}$ for passing energetic particles. (a) Co-passing branch $(\sigma=+1)$. (b) Counter-passing branch $(\sigma=-1)$, where $\sigma = \mathrm{sgn}(v_\parallel)$. Black curves indicate orbit boundaries and cyan points denote resonance branches obtained from Eq.~(\ref{eq:res_trapped}) for $|n|=6$ and $f=68\,\mathrm{kHz}$. The dashed grey curves correspond to the projected footprint in CoM space of the flux surface associated with the real-space mode location, and their proximity to the fluctuation ridge indicates that the dominant CoM response is linked to the same mid-radius region.}
    \label{fig:com_fluct_reso}
\end{figure}

The CoM distribution $F_{\mathrm{CoM}}(P_\varphi,\lambda)$ of the energetic particles at $T_{\mathrm{EP}} = 400$~keV, together with the orbit-topology boundaries, is shown in Appendix~\ref{app:com_distribution}.

The projection onto the dedicated EPCoM grid \cite{Brochard2024, BrochardEPCoM, Brochard2025NF} is defined on a structured grid in $(P_\varphi,\lambda,E)$ obtained from CoM-style mapping \cite{Brochard2024,Rud2024NF,BrochardEPCoM, Brochard2025NF}, given by

\begin{equation}
F_{\mathrm{CoM}}(P_\varphi,\lambda,E)
=
\int
F_{\mathrm{GYSELA}}(\mathbf{X},v_\parallel,\mu)\,
\delta^{(5)}\!\left[
\mathbf{Z}_{\mathrm{CoM}}(\mathbf{X},v_\parallel,\mu)
-
(P_\varphi,\lambda,E)
\right]
\frac{d\Gamma_{\mathbf{X},v_\parallel,\mu}}
     {d\Gamma_{P_\varphi,\lambda,E}},
\label{eq:CoM_Projection}
\end{equation}

where $\mathbf{Z}_{\mathrm{CoM}}$ denotes the
coordinate mapping from GYSELA phase space to CoM space and the 
Dirac delta enforces the coordinate mapping from 
$(\mathbf{X},v_\parallel,\mu)$ to $(P_\varphi,\lambda,E)$.
$d\Gamma_{\mathbf{Z}} = 2\pi \mathcal{J}_\mathrm{GYS} B_{\parallel s}^*\,d\psi\,d\theta\,
d\varphi\,dv_\parallel\,d\mu$ is the gyrocentre phase-space
volume element, and
$d\Gamma_{\mathrm{CoM}} = \mathcal{J}_\mathrm{CoM} dP_\varphi\,d\lambda\,dE$ is the
CoM space volume element with $\mathcal{J}_\mathrm{GYS}$ and $\mathcal{J}_\mathrm{CoM}$ representing the real space and CoM space Jacobian, respectively.
The separation of co- and counter-passing branches removes a degeneracy in mapping and makes resonant populations easier to interpret. In some regions of phase space, the same $(\Pphi, \lambda, E)$ corresponds to both a co-passing and a counter-passing orbit and the trapped response is split between the co- and counter-going parts to keep continuity across the trapped–passing boundary.
The black curves indicate orbit boundaries computed by EPCoM (see Appendix~\ref{app:com_distribution}), so the plot directly identifies which orbit classes populate which regions of $(P_\varphi,\lambda)$ space.

The fluctuation map $\tilde{F}_{\mathrm{CoM}}$ shown in Fig.\ref{fig:com_fluct_reso} quantifies wave-particle energy exchange in invariant space.
The fluctuation quantity is defined as the departure of the distribution function from its axisymmetric (toroidally averaged) mean. This definition is used here for the linear phase only.
The fluctuation is characterised at each phase-space point by the
root-mean-square of the departure from the toroidal mean,
\begin{equation}
\delta F_{\rm rms}(\mathbf{Z})
=\Big\langle \big[F_{EP}(\mathbf{X},v_\parallel,\mu)-\langle F_{EP}\rangle_\varphi(\mathbf{X},v_\parallel,\mu)\big]^2
\Big\rangle_\varphi^{1/2},
\qquad \mathbf{Z}=(r,\theta,v_\parallel,\mu),
\label{eq:dF_rms}
\end{equation}
which is projected onto CoM space as a phase-space-volume-weighted average,
\begin{equation}
\tilde F_{\rm CoM}(P_\varphi,\lambda,E)
=\frac{\displaystyle\int \delta F_{\rm rms}\,
\delta^{(3)}\!\left[\mathbf Z_{\rm CoM}(\mathbf Z)-(P_\varphi,\lambda,E)\right]
d\Gamma_{\mathbf Z}}
{\displaystyle\int
\delta^{(3)}\!\left[\mathbf Z_{\rm CoM}(\mathbf Z)-(P_\varphi,\lambda,E)\right]
d\Gamma_{\mathbf Z}},
\label{eq:deltaF_CoM}
\end{equation}
with $\mathbf Z$ as in \eqref{eq:dF_rms} and the mapping is that of
Eq.~(\ref{eq:CoM_Projection}). Each CoM bin therefore holds the orbit average of the local fluctuation amplitude.

Fig.\ref{fig:com_fluct_reso} shows the CoM fluctuation amplitude at target energy $E = 407$ keV, separated by orbit type into the co-passing ($\sigma = +1$,  Fig.\ref{fig:com_fluct_co}) and counter-passing ($\sigma = -1$, Fig.\ref{fig:com_fluct_counter}) branches.
In both passing branches a clear fluctuation ridge is visible, aligned with the resonance curves (cyan points) computed from Eq.~(\ref{eq:res_trapped}), for $|n|=6$ this gives $\omega_0 = 4.3\times10^{5}\,\mathrm{rad/s}$ ($f=68$ kHz).
Overlaying the resonance curves on the distribution of the CoM fluctuations directly identifies the orbit families responsible for the drive. The TAE growth rate is set by \big[$n\partial F/\partial \Pphi + \omega\big(\partial F/\partial E - (\lambda/E)\partial F/\partial \lambda\big)$\big] evaluated at the resonance, and Fig.\ref{fig:com_fluct_reso} shows that mostly the deeply co-passing orbits at target energy $E=407$ keV intersect the steepest gradient region. The counter-passing branch shows a comparatively weaker response, identifying co-passing particles as the dominant contributors to the TAE drive at this energy.

In addition, the dashed grey curves corresponding to the flux surface near the real-space mode location pass close to the fluctuation ridge in CoM space. For the present ITPA-TAE case, the mode is localised around $r/a \approx 0.5$, which corresponds roughly to $\psi/\psi_{\mathrm{edge}} \sim 0.25$ under the usual $\psi \propto r^2$ mapping, where $\psi_{\mathrm{edge}}$ is the poloidal flux at the last closed flux surface. When this flux surface is projected into $(P_\varphi,\lambda)$ space, its footprint is found to lie close to the region where the CoM fluctuation is strongest. This gives a simple and intuitive link between the real-space mode localisation and the invariant-space response. Although the two are not expected to overlap point by point, their close proximity suggests that the dominant fluctuation is associated with orbit classes that access the same mid-radius region.

\subsection{NLED-AUG on-axis benchmark}

\begin{figure}[!htbp]
    \centering
    \includegraphics[width=\textwidth]{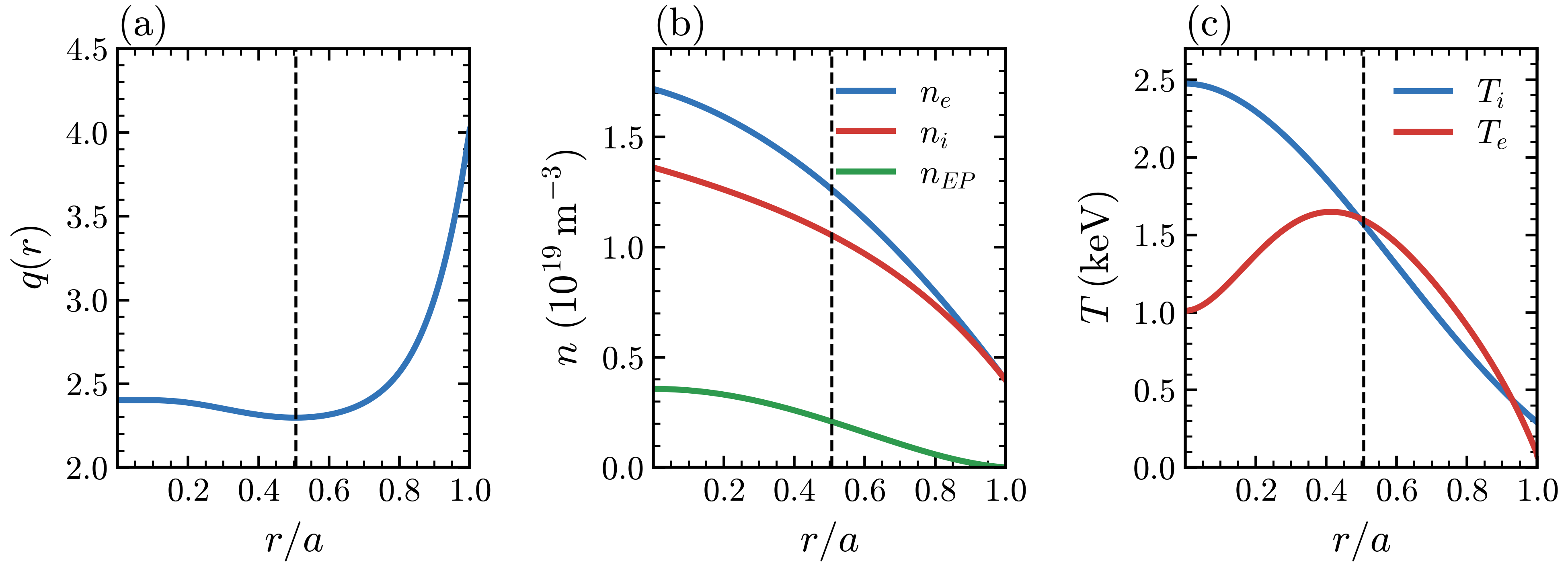}
    \caption{NLED-AUG on-axis benchmark:
    (a) reversed-shear safety factor profile used in GYSELA.
    (b) Input density profiles (electrons fixed, fast-ion density scaled,
    quasineutrality enforced through $n_i = n_e - n_{\mathrm{EP}}$).
    (c) Temperature profiles $T_i(r)$ and $T_e(r)$ and
    EP temperature treated as a scan parameter.
    The dashed line marks the position of $q_{\min}$.}
    \label{fig:nled_safety_profiles}
\end{figure}

The NLED-AUG on-axis benchmark provides a more
stringent validation than the ITPA-TAE case, owing
to its realistic reversed-shear safety factor profile
and high energetic-particle pressure.
The reference equilibrium used by HYMAGYC, MEGA,
and ORB5 is based on the ASDEX-U discharge \#31213 at
$t = 0.84\,\mathrm{s}$, reconstructed with CHEASE
from the experimental EQDSK file~\cite{Vlad2021NLED}.
In the present GYSELA simulations, 
CULHAM equilibrium \cite{Connor1988Culham} is used instead, which reproduces
the key features of the NLED-AUG configuration, a
reversed-shear safety factor profile with
$q_{\min}\approx2.30$, elongation $\kappa=1.8$,
and aspect ratio $R/a=2.57$. The analytical equilibrium shifts the continuum accumulation frequency with respect to the reconstructed equilibrium, which explains most of the frequency offset discussed later in this section. Still, it retains the reversed-shear safety factor profile and the local gradients that
set the energetic-particle drive.
The equilibrium profiles used in GYSELA are shown in Fig.~\ref{fig:nled_safety_profiles}
and follow the NLED-AUG benchmark specification~\cite{Vlad2021NLED}.
The minimum of the safety factor is $q_{\min}\simeq2.30$ at $r/a\simeq0.51$, about $1\%$
above the reference value $q_{\min}\approx2.28$.
The safety factor profile, shown in Fig.~\ref{fig:nled_safety_profiles}(a), produces a
non-monotonic Alfv\'en continuum that supports RSAE-type modes near $q_{\min}$,
consistent with the benchmark configuration~\cite{Vlad2021NLED}. 

The RSAE frequency and location are set mainly by $q_{\min}$ and the local Alfv\'en speed~\cite{Breizman2005,Kramer2004}.
The ion and electron temperature profiles are those of the benchmark specification
(Fig.~\ref{fig:nled_safety_profiles}(c)). The electron density is held fixed, while the bulk
ion density is obtained self-consistently from
quasi-neutrality $n_i = n_e - n_{EP}$~\cite{Vlad2021NLED}.
The EP species consists of deuterons ($A_{EP}=2$,
$Z_{EP}=1$) with a peaked on-axis density profile,
initialised with the canonical Maxwellian
(Sec.~\ref{sec:canonical_maxwellian}).
For the reference case, the EP-to-bulk-ion density
ratio is $n_{EP}/n_{i0} = 0.261$, where $n_{i0}$ is the
on-axis bulk ion density, and the EP
temperature is $T_{EP} = 93\,\mathrm{keV}$, constant
in radius, consistent with the NLED-AUG benchmark
definition~\cite{Vlad2021NLED}.

\paragraph{Numerical parameters.}
The NLED-AUG simulations are performed on a
five-dimensional grid
$(N_r,N_\theta,N_\varphi,N_{v_\parallel},N_\mu)
= (255,128,32,128,63)$
on a full torus ($\Delta\varphi = 2\pi$), with
aspect ratio $R/a = 2.57$ and $a=252.50$ $\rho_i$ in the parameters of CULHAM geometry.
The magnetic equilibrium uses the analytical CULHAM
geometry with elongation $\kappa = 1.8$, and the
safety factor profile is prescribed analytically,
consistent with the reversed-shear NLED-AUG
configuration~\cite{Vlad2021NLED}.
A single toroidal mode number $n=1$ is retained,
with the dominant poloidal harmonic $m=2$.
Three plasma species, thermal ions, 
fully kinetic electrons, and energetic particles are evolved simultaneously in the collisionless regime. 

\begin{figure}[!htbp]
    \centering
    \begin{subfigure}[b]{0.45\textwidth}
        \centering
        \includegraphics[width=\linewidth]{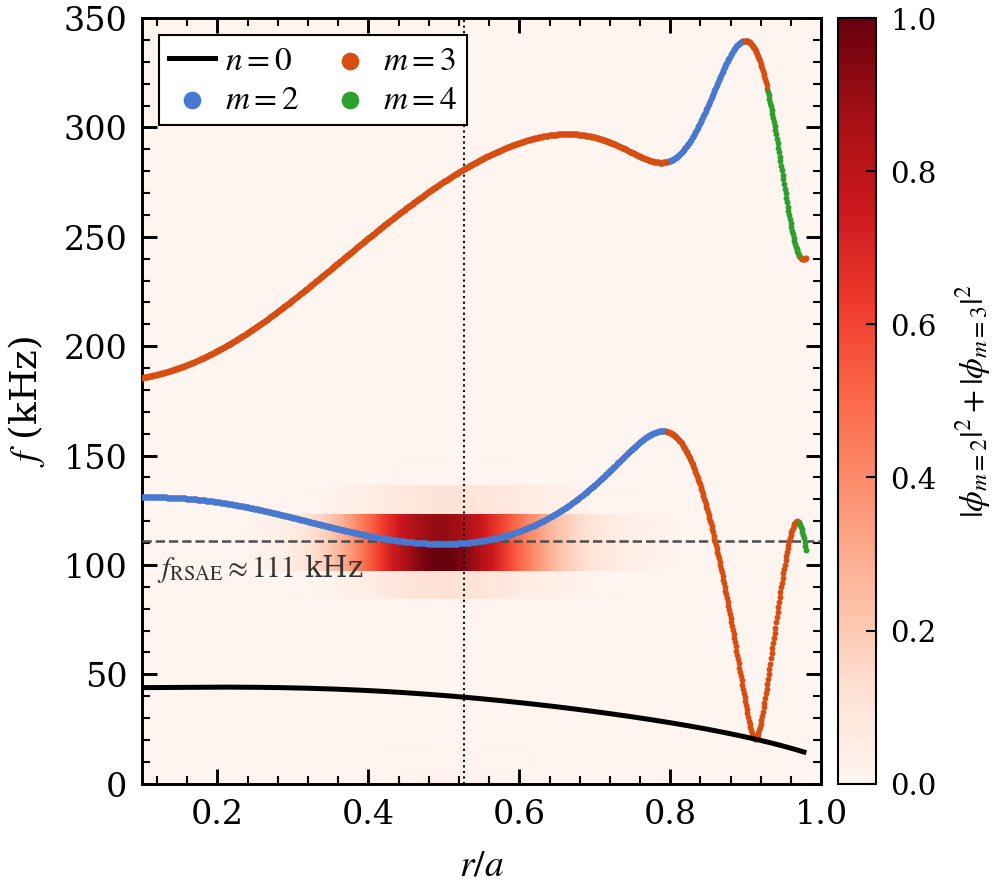}
        \caption{}
        \label{fig:nled_continuum_m3}
    \end{subfigure}
    \hfill
    \begin{subfigure}[b]{0.49\textwidth}
        \centering
        \includegraphics[width=\linewidth]{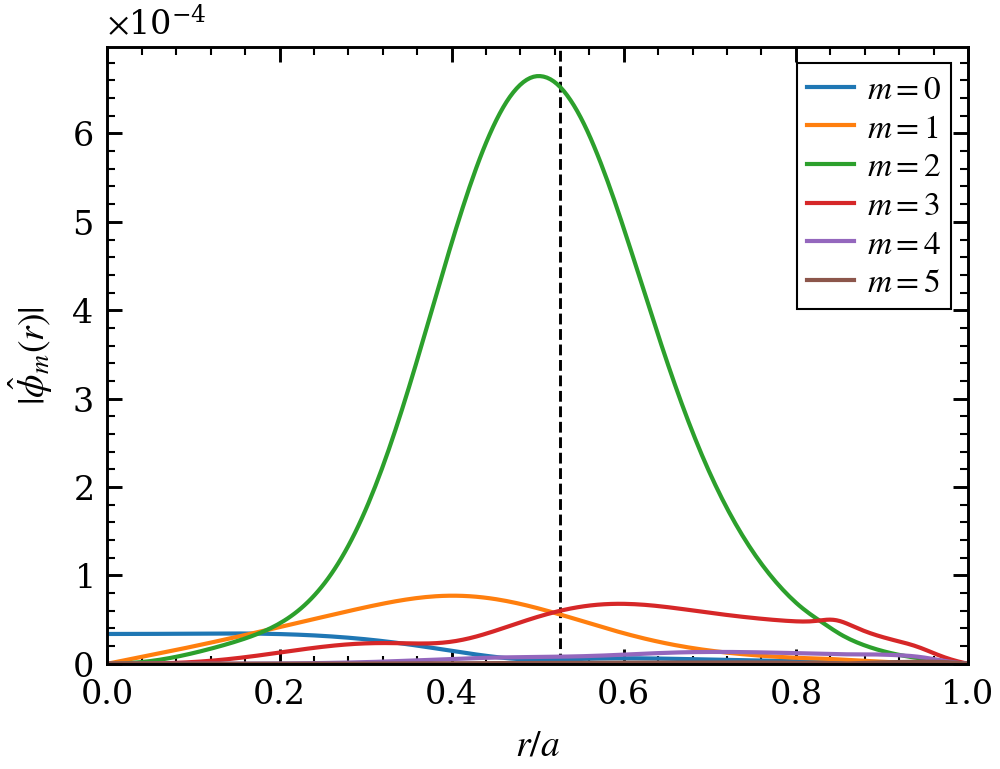}
        \caption{}
        \label{fig:nled_fft}
    \end{subfigure}
    
    \vspace{1em}
    
    \begin{subfigure}[b]{0.49\textwidth}
        \centering
        \includegraphics[width=\linewidth]{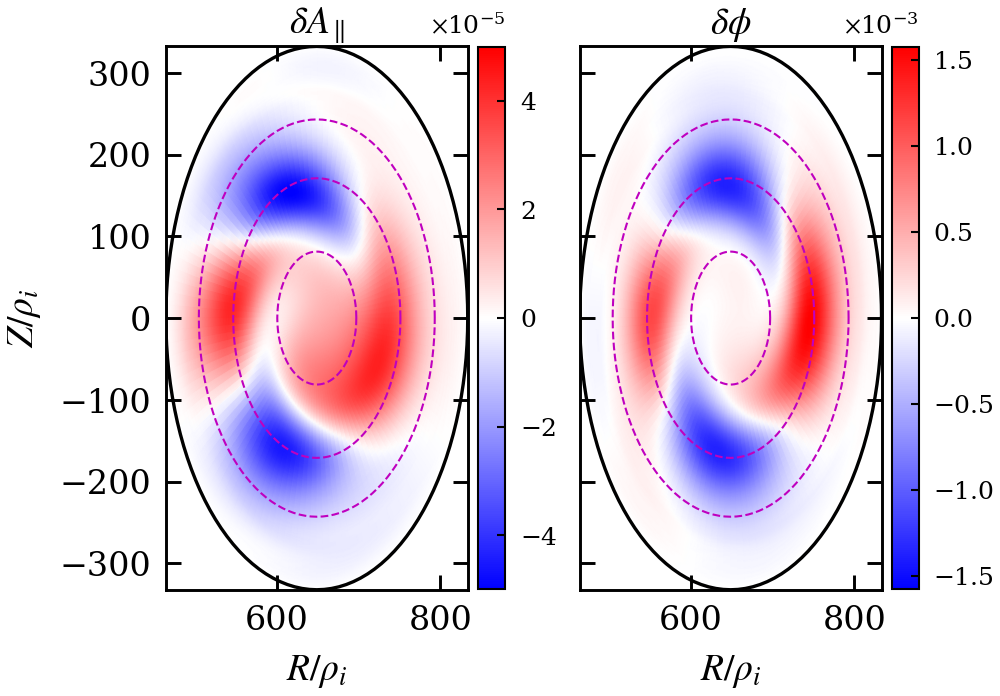}
        \caption{}
        \label{fig:nled_phi2d}
    \end{subfigure}
    \hfill
    \begin{subfigure}[b]{0.49\textwidth}
        \centering
    \end{subfigure}
    
    \caption{NLED-AUG on-axis benchmark.
    (a) Shear Alfv\'en continuum for $n=1$ computed with ALCON~\cite{Deng2012NF}
    from the GYSELA CULHAM equilibrium ($m=2$ blue, $m=3$ orange, $m=4$ green;
    black: local geodesic acoustic frequency from the $n=0$ continuum),
    with the normalised spectral power $|\hat\phi_{m=2}|^2+|\hat\phi_{m=3}|^2$
    as a heatmap.
    (b) Radial profiles of the poloidal harmonics $|\hat\phi_{m}(r)|$ in the
    straight-field-line angle; the dashed line marks $q_{\min}$.
    (c) Poloidal cross-sections of $A_\parallel$ and $\phi$ during the linear phase.}
    \label{fig:nled_mode_structure}
 
\end{figure}

\paragraph{Alfv\'en continuum and mode identification.}
The Alfv\'en continuum structure for the NLED-AUG
case is shown in Fig.\ref{fig:nled_continuum_m3},
computed with the ideal-MHD continuum solver
ALCON~\cite{Deng2012NF} using the magnetic
equilibrium generated by GYSELA.
For $n=1$, the relevant poloidal harmonics are
$m=2$ and $m=3$, shown together with $m=4$ for
reference.
The reversed-shear safety factor profile, with
$q_{\min}\approx2.30$ at $r/a\approx0.51$,
shows a non-monotonic structure on the $m=2$
branch. The branch starts near $f\approx130\,\mathrm{kHz}$
at the innermost radius shown $(r/a \approx 0.1)$, the branch decreases to a local
minimum before rising again
towards the edge.
This minimum constitutes the continuum accumulation
point characteristic of reversed-shear equilibria.
The $m=3$ branch runs significantly higher, spanning
$185$--$340\,\mathrm{kHz}$ across the plasma radius.
Both branches drop sharply towards zero near
$r/a\approx0.92$, where the safety factor rises
steeply towards the edge.
As illustrated in Fig.\ref{fig:nled_continuum_m3}, the combined spectral power of the dominant poloidal harmonics ($\vert{}\phi_{m=2}\vert{}^2 + \vert{}\phi_{m=3}\vert{}^2$) exhibits a strong localised maximum concentrated right at the local extremum (minimum) of the $m=2$ continuous shear Alfvén branch near $r/a \approx 0.5$. This pronounced peak at an accumulation frequency of approximately $f_{\text{RSAE}} \approx 111\text{ kHz}$ demonstrates that the observed discrete eigenmode is indeed an RSAE localised near the minimum safety factor surface. It is to be noted here that the black curve in Fig.\ref{fig:nled_continuum_m3} is the local geodesic acoustic
frequency obtained from the $n = 0$ continuum. It forms the floor of the shear Alfv\'en
continuum at rational surfaces: at the $q = 3$ surface ($r/a \simeq 0.91$) the $m = 3$
branch terminates on it at $20.6$~kHz, and the beta-induced Alfv\'en eigenmode (BAE) gap
lies below it. At the mode location this floor is $\simeq 40$~kHz, whereas the observed mode
at $f \simeq 111$~kHz lies well above it, at the accumulation point of the $m = 2$ branch.
The mode frequency is therefore set by the shear-Alfv\'en dispersion rather than by
geodesic compressibility, hence we identify the mode as an RSAE mode. No spectral power is observed in the TAE gap.

\paragraph{Reference case and multi-code comparison.}
Table~\ref{tab:nled_comparison} compares the GYSELA
eigenfrequency, growth rate, and drive sensitivity
against HYMAGYC, MEGA, and
ORB5~\cite{Vlad2021NLED} at the reference parameters
($n_{EP}/n_{i0} = 0.261$, $T_{EP} = 93\,\mathrm{keV}$).
HYMAGYC and MEGA both identify the RSAE as the
dominant mode, with frequencies
$\omega\approx-127$ and $-134\,\mathrm{kHz}$,
respectively.
GYSELA obtains $\omega=-111\,\mathrm{kHz}$,
consistent with the RSAE branch but shifted
slightly towards lower absolute frequency. This offset likely reflects the use of the analytical CULHAM equilibrium in place of the CHEASE-reconstructed EQDSK, to which the RSAE frequency is very sensitive through $q_{min}$.

ORB5 observes an externally localised TAE
($\omega\approx-168\,\mathrm{kHz}$) as the dominant
mode at the reference density and the RSAE becomes
dominant only for
$n_{EP}/n_{i0}\gtrsim0.4$~\cite{Vlad2021NLED}.
GYSELA does not observe this external TAE either.

There are two likely reasons for the absence of TAEs.
First, in the CULHAM analytical geometry, the external TAE is radially localised near $r/a\approx0.7$--$0.9$,
close to the SOL boundary, where the absorbing
boundary condition introduces additional numerical
damping.
Second, the canonical Maxwellian distribution is different from the local Maxwellian used by other codes.

The drive sensitivity, quantified by the growth rate
slopes $S_n \equiv \partial\gamma/\partial(n_{EP}/n_{i0})$ and
$S_T \equiv \partial\gamma/\partial T_{EP}$, shows that the temperature slope $S_{T}$ lies between the reference values, while the density slope $S_{n}$ is lower than both for GYSELA,
as summarised in Table~\ref{tab:nled_comparison}, the
individual scans are discussed below.

\begin{table}[htbp]
\centering
\caption{NLED-AUG on-axis benchmark: RSAE
characteristics at the benchmark EP parameters
$n_{EP}/n_{i0} = 0.261$, $T_{EP} = 93\,\mathrm{keV}$,
and drive sensitivity from parameter scans.
$S_n \equiv \partial\gamma/\partial(n_{EP}/n_{i0})$
and $S_T \equiv \partial\gamma/\partial T_{EP}$ are
the linear drive sensitivities with respect to EP
density fraction and temperature, respectively,
obtained from linear fits to the scan data.
Reference values digitised from~\cite{Vlad2021NLED}. $^\dagger$ORB5 observes an externally localised TAE at the benchmark density and the RSAE
becomes dominant for $n_{EP}/n_{i0}\gtrsim0.4$~\cite{Vlad2021NLED} and therefore no RSAE slopes are
given.}
\label{tab:nled_comparison}
\setlength{\tabcolsep}{4pt}
\begin{tabular*}{\linewidth}{@{\extracolsep{\fill}}lccccc@{}}
\toprule
Code
  & $\omega$
  & $\gamma$
  & Mode
  & $S_n$
  & $S_T$ \\
  & (kHz)
  & ($10^4\,\mathrm{s}^{-1}$)
  &
  & ($10^5\,\mathrm{s}^{-1}$)
  & ($10^6\,\mathrm{s}^{-1}\,\mathrm{MeV}^{-1}$) \\
\midrule
GYSELA       & $-111$ & $5.6$ & RSAE          & $2.7$ & $2.4$ \\
HYMAGYC      & $-127$ & $8.0$ & RSAE          & $5.3$ & $3.1$ \\
MEGA         & $-134$ & $2.7$ & RSAE          & $3.5$ & $2.1$ \\
ORB5         & $-168$ & $11.5^\dagger$
                              & TAE$^\dagger$ & --    & --    \\
\bottomrule
\end{tabular*}
\end{table}

\begin{figure}[!htbp]
    \centering
    \begin{subfigure}[b]{0.48\textwidth}
        \centering
        \includegraphics[width=\linewidth]{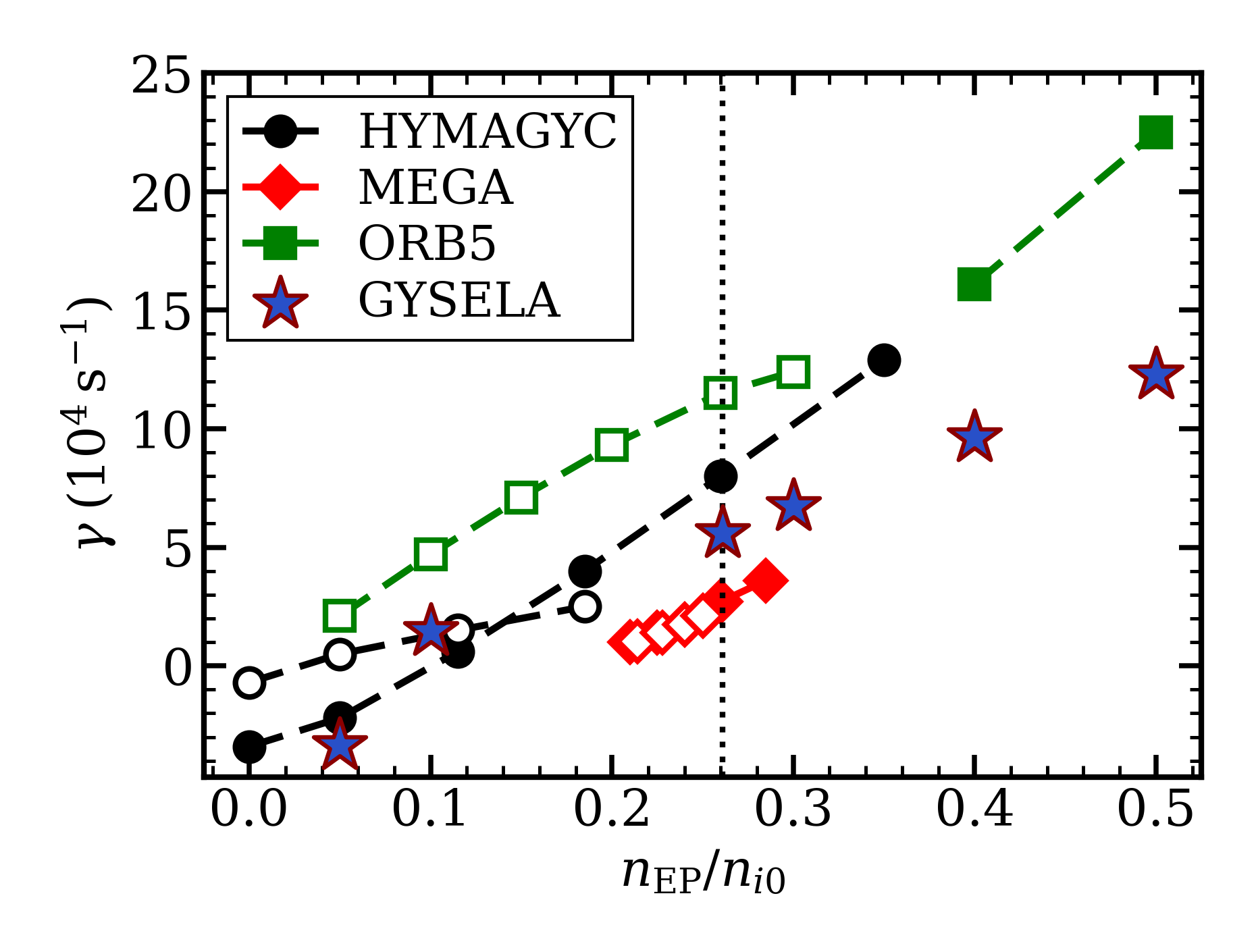}
        \caption{}
        \label{fig:nled_gamma_frac}
    \end{subfigure}
    \hfill
    \begin{subfigure}[b]{0.48\textwidth}
        \centering
        \includegraphics[width=\linewidth]{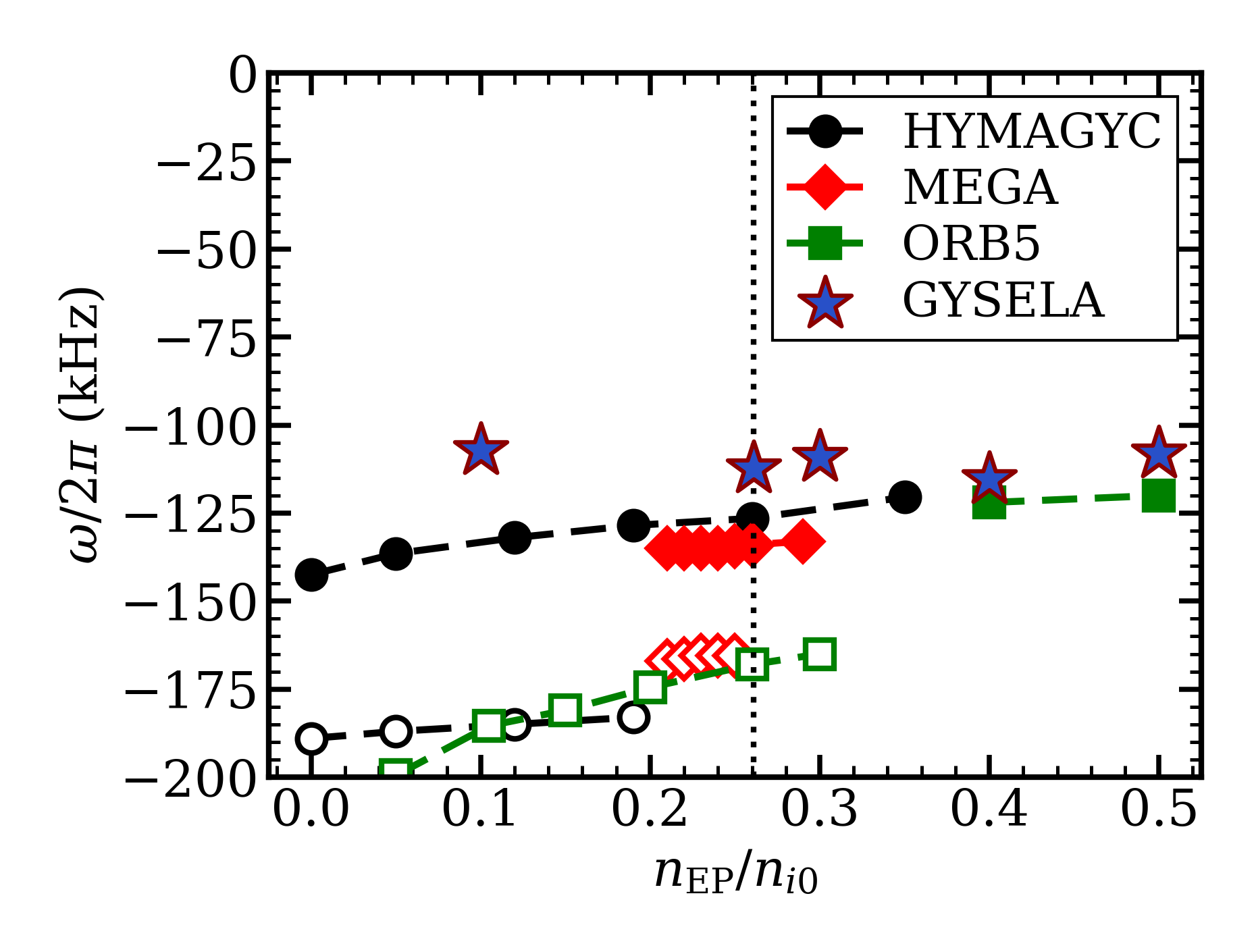}
        \caption{}
        \label{fig:nled_omega_frac}
    \end{subfigure}

    \vspace{0.3cm}

    \begin{subfigure}[b]{0.48\textwidth}
        \centering
        \includegraphics[width=\linewidth]{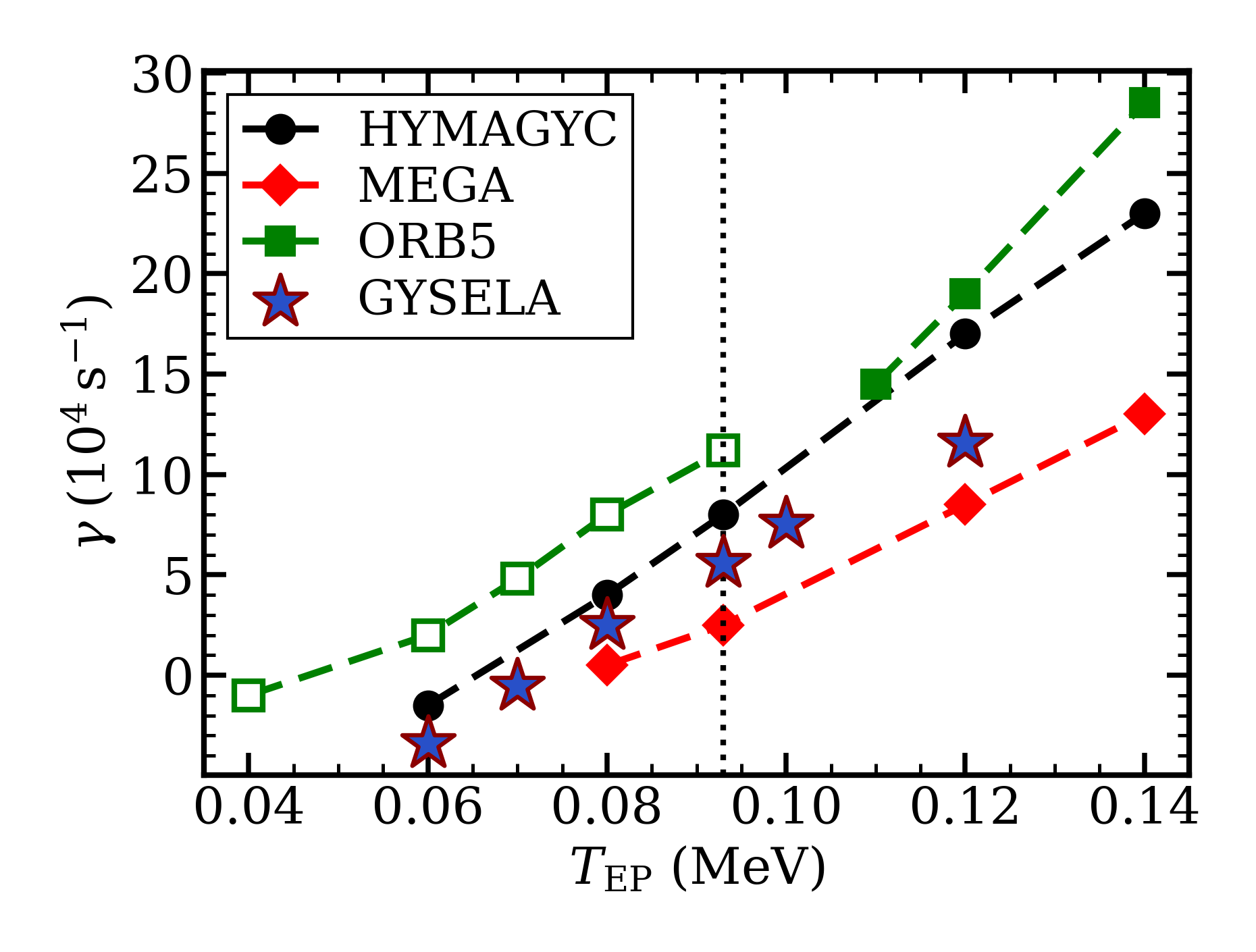}
        \caption{}
        \label{fig:nled_gamma_temp}
    \end{subfigure}
    \hfill
    \begin{subfigure}[b]{0.48\textwidth}
        \centering
        \includegraphics[width=\linewidth]{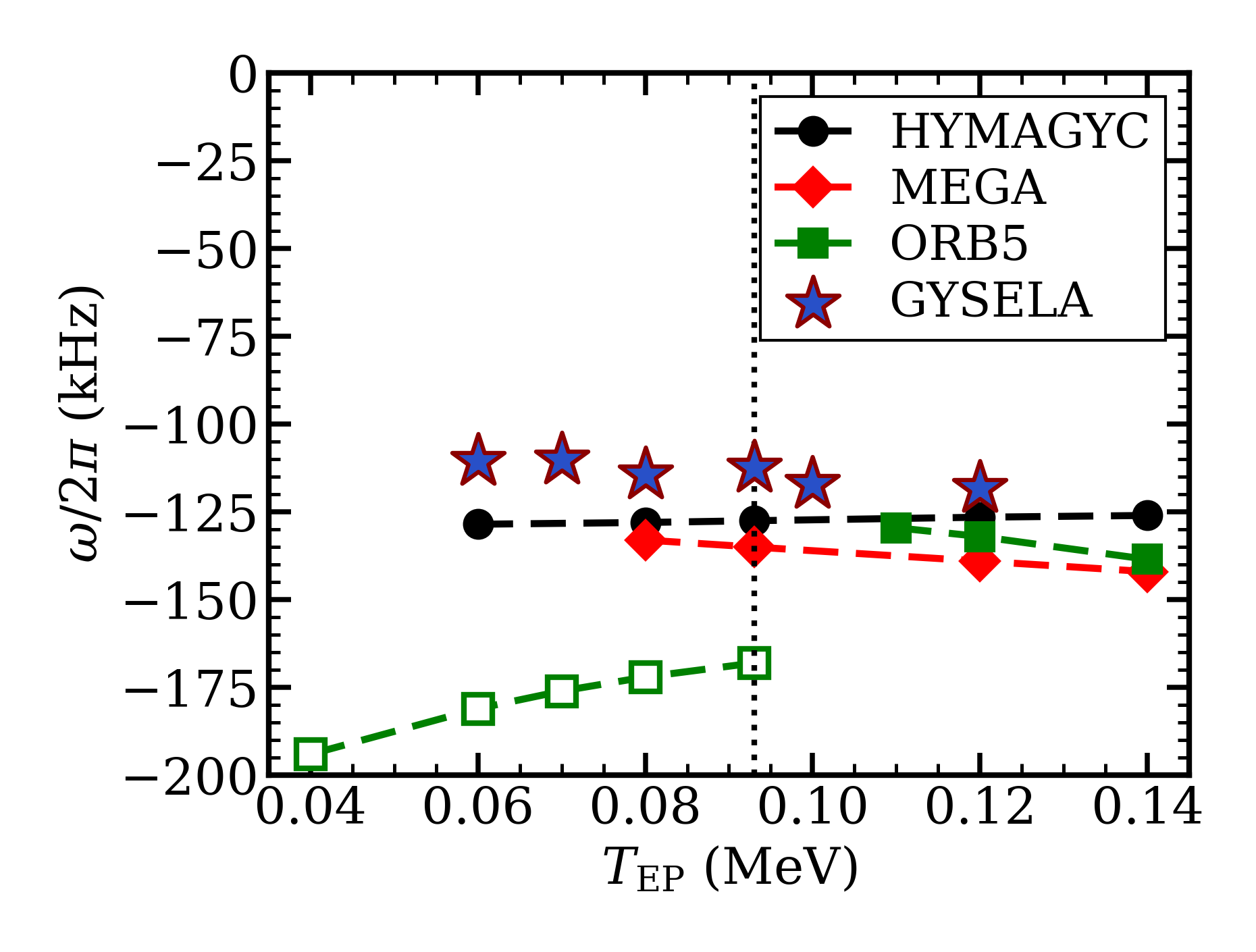}
        \caption{}
        \label{fig:nled_omega_temp}
    \end{subfigure}

    \caption{NLED-AUG on-axis benchmark.
    (a) Growth rate $\gamma$ and (b) real frequency $\omega$ versus the on-axis
    fast-ion fraction $n_{\mathrm{EP}}/n_{i0}$ at fixed $T_{\mathrm{EP}} = 93\,\mathrm{keV}$,
    the dashed vertical line marks the reference value $n_{\mathrm{EP}}/n_{i} = 0.261$.
    (c) Growth rate $\gamma$ and (d) real frequency $\omega$ versus the fast-ion
    temperature $T_{\mathrm{EP}}$ at fixed $n_{\mathrm{EP}}/n_{i0} = 0.261$.
    Markers: GYSELA, curves: HYMAGYC, MEGA and ORB5 results
    from Ref.~\cite{Vlad2021NLED}, with closed symbols for the RSAE and open
    symbols for the TAE. The GYSELA scans were performed with a monotonic $T_e$ profile, and at the reference parameters this changes the RSAE frequency by less than $1\%$.}
    \label{fig:nled_scans}
\end{figure}

\paragraph{Energetic-particle density scan.}
Fig.~\ref{fig:nled_gamma_frac} shows the linear growth rate $\gamma$ as a function of the
EP density fraction $n_{EP}/n_{i0}$, with $T_{EP}=93\,\mathrm{keV}$ fixed.
The GYSELA growth rate crosses zero near $n_{EP}/n_{i0}\approx0.08$, indicating the
threshold EP density below which the RSAE is stable.
Above this threshold, $\gamma$ increases monotonically with EP fraction, consistent with
stronger phase-space gradients providing greater drive.
The resulting slope $S_n$ is lower than both the MEGA and HYMAGYC values
(Table~\ref{tab:nled_comparison}). In Ref.~\cite{Vlad2021NLED}, the differences in the
threshold among the reference codes are attributed mainly to the different damping
included in each code. The agreement in growth rate for this case is therefore
qualitative rather than quantitative, as expected for an RSAE, which is very sensitive to
the equilibrium and the EP phase-space distribution.

The real frequency $\omega$ as a function of density fraction is shown in
Fig.~\ref{fig:nled_omega_frac}. GYSELA shows a weak, non-monotonic variation across the
scan, consistent with the RSAE frequency being governed primarily by the continuum
accumulation point rather than by the EP drive intensity. HYMAGYC shows a somewhat larger
variation, while MEGA remains nearly flat. Despite the offset in absolute frequency, all
codes agree that $\omega$ is weakly sensitive to EP density, confirming the RSAE character
of the observed mode.

\paragraph{Energetic-particle temperature scan.}
Fig.~\ref{fig:nled_gamma_temp} shows the growth rate as a function of EP temperature
$T_{EP}$, with $n_{EP}/n_{i0}=0.261$ fixed.
GYSELA shows negative growth rates for $T_{EP}\lesssim75\,\mathrm{keV}$, indicating a
temperature threshold for RSAE instability.
A similar threshold is found in HYMAGYC (unstable near $T_{EP}\approx70\,\mathrm{keV}$),
while MEGA is marginally unstable at $80\,\mathrm{keV}$, its lowest scanned value.
Above threshold, $\gamma$ increases with $T_{EP}$ in all codes, and the slope $S_T$
obtained by GYSELA falls between the MEGA and HYMAGYC values.
The real frequency shown in Fig.~\ref{fig:nled_omega_temp} varies weakly with $T_{EP}$,
remaining close to $-111\,\mathrm{kHz}$ in GYSELA (with the sign convention of
Ref.~\cite{Vlad2021NLED}, i.e. positive in the electron diamagnetic direction),
$-127\,\mathrm{kHz}$ in HYMAGYC and $-134\,\mathrm{kHz}$ in MEGA across the whole scan.
This confirms that the mode frequency is pinned near the continuum accumulation point and
is insensitive to the EP temperature.
Together, the density and temperature scans show that GYSELA reproduces the parametric
dependence of the RSAE drive found by the reference codes in a reversed-shear geometry.

\paragraph{Mode spatial structure.}
The radial structure of the dominant poloidal
harmonics is shown in Fig.\ref{fig:nled_fft}.
The mode is localised near the $q_{\min}$ surface
at $r/a\approx0.5$, with $m=2$ as the dominant
harmonic, reflecting the local safety factor
$q\approx2.30$ at the mode location.
Two-dimensional poloidal cross-sections of
magnetic vector potential $A_\parallel$ and electric potential $\phi$ are presented in
Fig.\ref{fig:nled_phi2d}, confirming the expected
Alfv\'enic spatial pattern and radial localisation
near the reversed-shear region.
Agreement in eigenfrequency, growth rate, spatial
structure, and parametric trends ensures that the
benchmark match extends well beyond scalar
quantities.

To map the wave--particle energy exchange in phase space and attribute the drive to orbit
classes, the local power transfer ($P=-\mathbf J\cdot\mathbf E$) between the
energetic particles and the $n\neq0$ fields is evaluated at every point
of the five-dimensional grid from the saved distribution and fields,
using the gyrocentre velocities of Eqs.~\eqref{eq:motion_gyrocenter_position}-- \eqref{eq:motion_gyrocenter_velocity},
\begin{equation}
P=-\int d\Gamma_{\mathbf Z}\,\bar F_{EP}\;
\dot{\mathbf X}\cdot\mathbf E_{\rm gc},\qquad
\mathbf E_{\rm gc}=-\nabla\mathcal J[\Phi]-\mathbf b\,\partial_t\mathcal J[A_\parallel],
\label{eq:power}
\end{equation}
where $\dot{\mathbf X}$ contains the parallel motion along $\mathbf b^*$,
the magnetic drift and the $A_\parallel$-flutter velocity, and the $n=0$
contribution is removed by repeating the evaluation with the toroidally
averaged fields. The gyroaverage $\mathcal J$ and the phase-space measure
are those of Eq.~\eqref{eq:Astar} and Eq.~\eqref{eq:volume_gyrocenter}. Particles are classified as trapped when
$|v_\parallel|<\sqrt{2\mu\,(B_{\max}(r)-B)/m_{\rm EP}}$, with
$B_{\max}(r)=\max_{\theta}B(r,\theta)$, and the remaining passing particles
are classified as co- or counter-passing according to the sign of
$v_\parallel$ ($\sigma=+1$ and $\sigma=-1$, respectively), with $v_\parallel>0$
parallel to the plasma current.

This local criterion ignores the finite orbit width effect and therefore is an approximate estimation of the drive. The power P of Eq.~\eqref{eq:power} is evaluated in GYSELA normalised units and below, the contribution of each orbit class is given as a fraction of the net power exchanged between the energetic particles and the mode. The counter-passing energetic particles drive the RSAE ($+1.59$), while
co-passing particles absorb energy from it ($-0.70$), about $44\%$ of
the counter-passing drive, the trapped particles add a small drive
($+0.11$).

\begin{figure}[!htbp]
    \centering
    \begin{subfigure}[b]{0.49\textwidth}
        \centering
        \includegraphics[width=\linewidth]{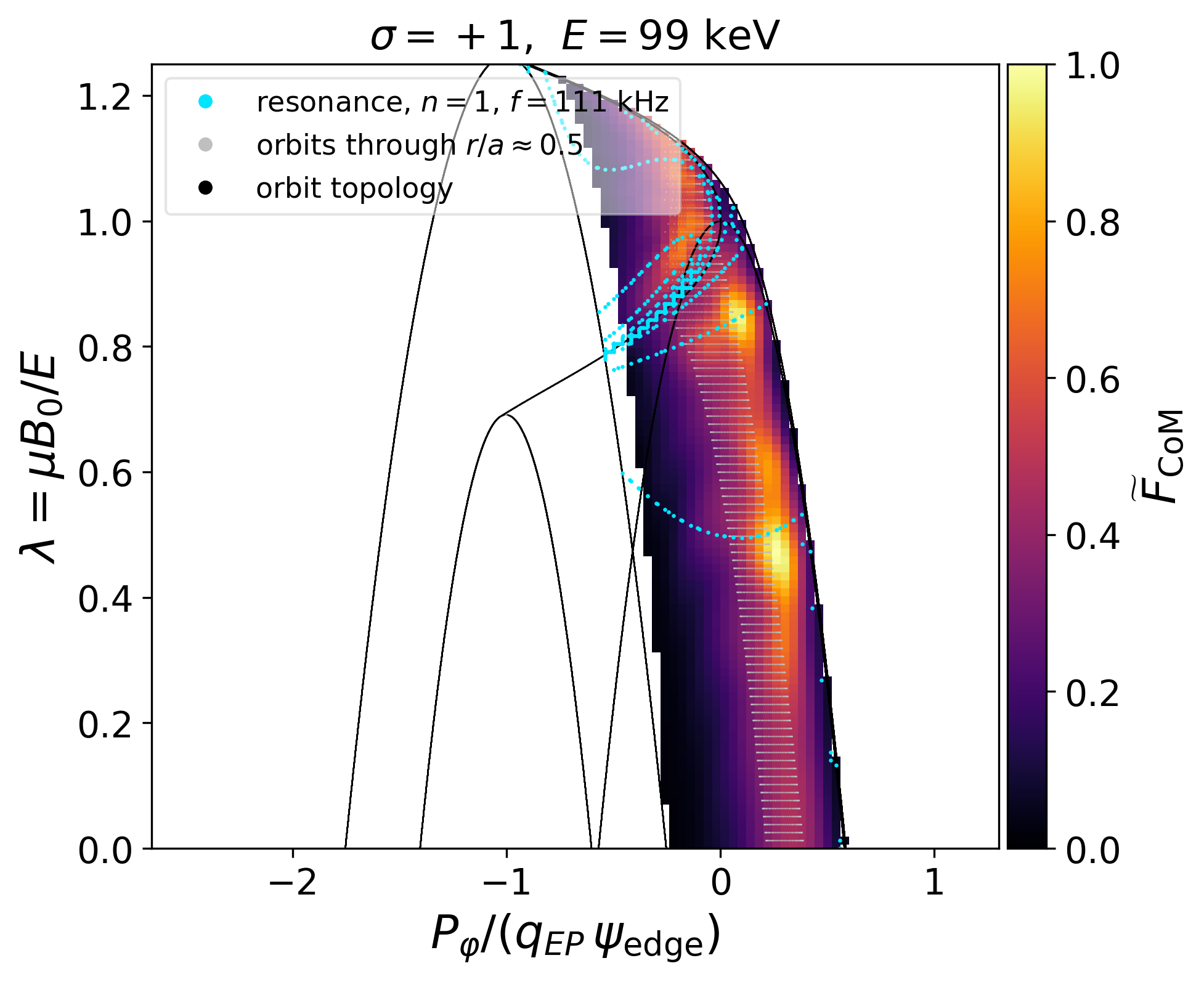}
        \caption{$\sigma=+1$, $E=99$~keV}
        \label{fig:nled_com_co99}
    \end{subfigure}
    \hfill
    \begin{subfigure}[b]{0.49\textwidth}
        \centering
        \includegraphics[width=\linewidth]{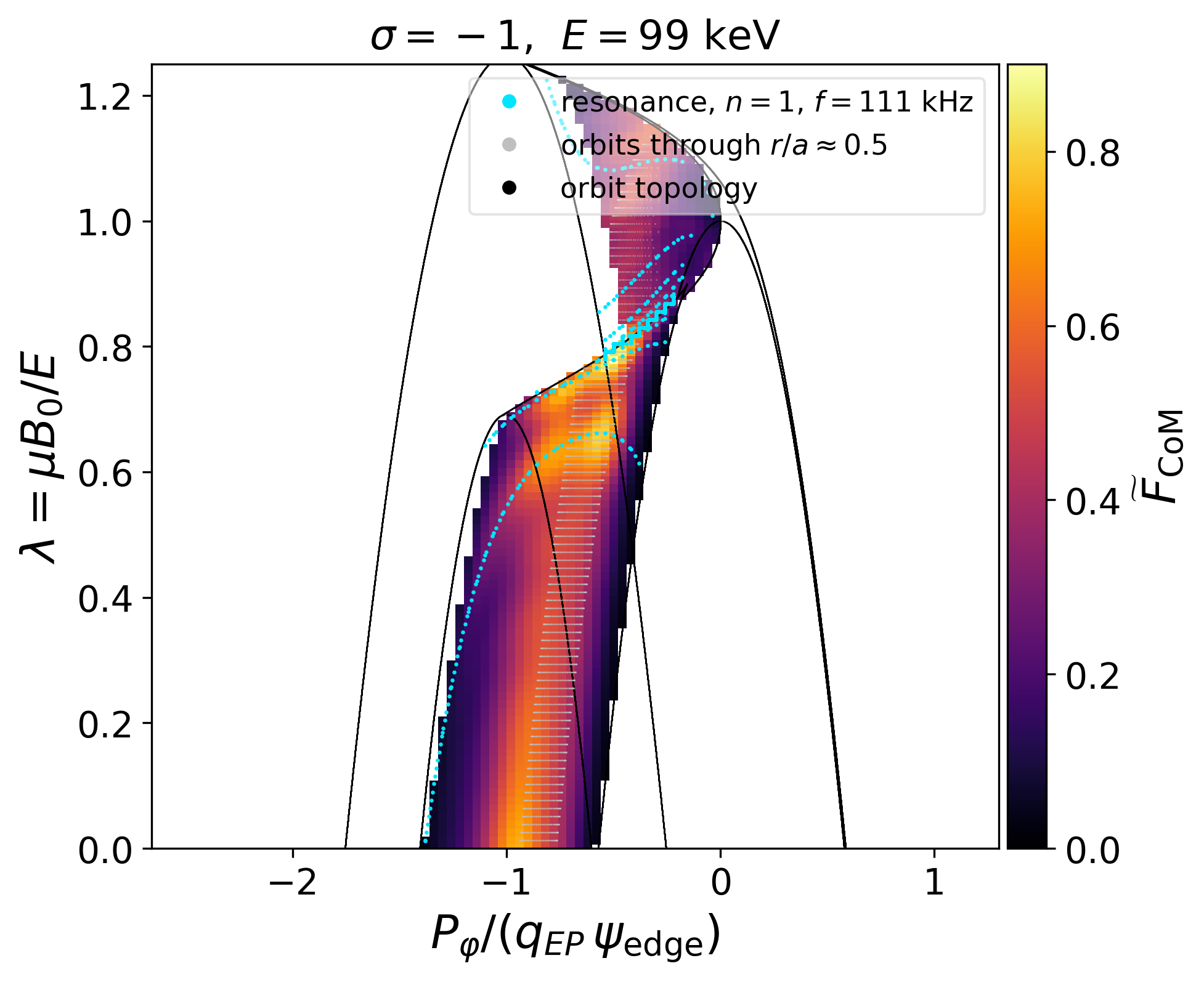}
        \caption{$\sigma=-1$, $E=99$~keV}
        \label{fig:nled_com_counter99}
    \end{subfigure}

    \begin{subfigure}[b]{0.49\textwidth}
        \centering
        \includegraphics[width=\linewidth]{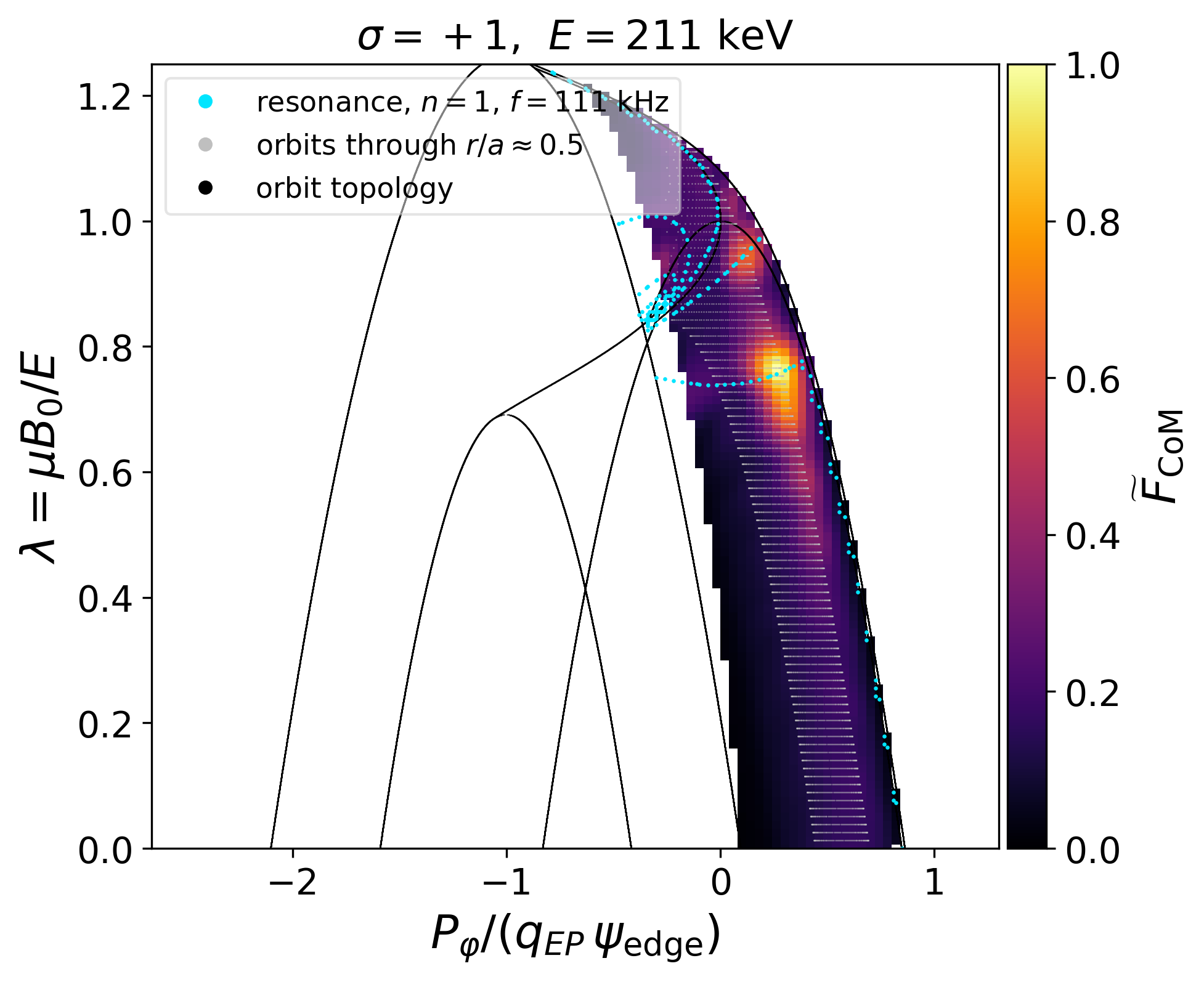}
        \caption{$\sigma=+1$, $E=211$~keV}
        \label{fig:nled_com_co211}
    \end{subfigure}
    \hfill
    \begin{subfigure}[b]{0.49\textwidth}
        \centering
        \includegraphics[width=\linewidth]{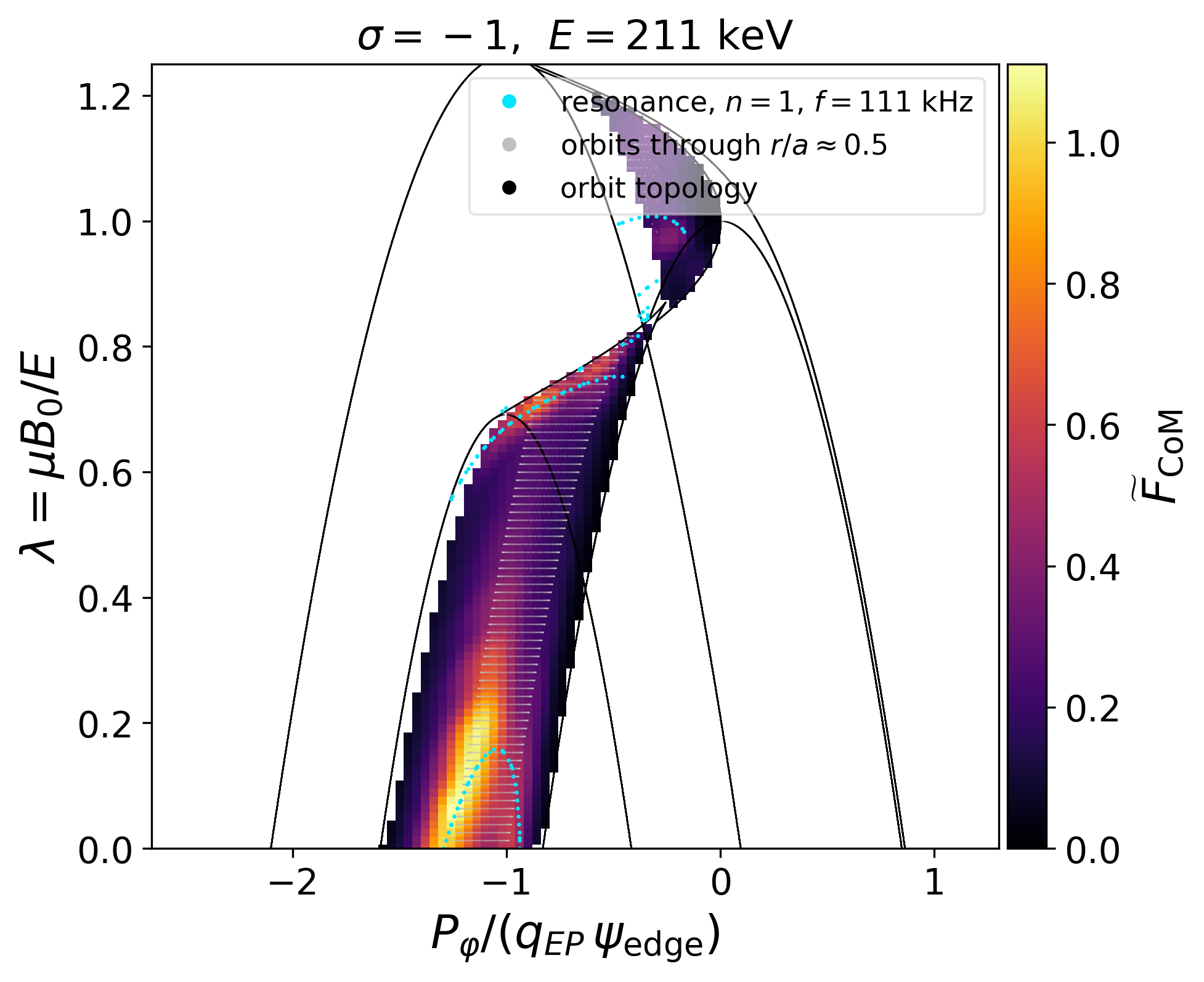}
        \caption{$\sigma=-1$, $E=211$~keV}
        \label{fig:nled_com_counter211}
    \end{subfigure}
    \caption{Normalised EP fluctuation $\widetilde F_{\rm CoM}(P_\varphi,\lambda)$ for the
    NLED-AUG case at $E=99$~keV (a, b) and $E=211$~keV (c, d), co-passing ($\sigma=+1$)
    and counter-passing ($\sigma=-1$). In each row both panels are normalised to the
    co-passing maximum (Eq.~\eqref{eq:deltaF_CoM}). Cyan dots: $n=1$ resonances of
    Eq.~\eqref{eq:res_trapped} at $f=111$~kHz; grey dots:
    orbits reaching $r/a\simeq0.5$; black dots: orbit-topology boundaries.}
    \label{fig:nled_com}
\end{figure}

The CoM projection of the fluctuation shows, directly from the distribution function,
which orbits are perturbed by the mode, and thus provides an independent check of the
resonant populations. Fig.~\ref{fig:nled_com} shows $\widetilde F_{\rm CoM}(P_\varphi,\lambda)$ at $E=99$~keV,
close to $T_{EP}=93$~keV, and at $E=211$~keV. The higher energy is chosen where the
counter-passing drive, summed over $(P_\varphi,\lambda)$, is largest, while at
$E\simeq T_{EP}$ the co-passing damping and the counter-passing drive nearly cancel. In each row both panels use the same colour scale, normalised to the maximum of the co-passing panel. The $n=1$
resonance lines of Eq.~\eqref{eq:res_trapped} at $f_{\rm RSAE}\simeq111$~kHz are overlaid
for a range of bounce/transit harmonics $p$, together with the orbits that reach
$r/a\simeq0.5$ and the orbit-topology boundaries. The fluctuation amplitude shows which
orbits interact with the mode, but not the sign of the exchange, which is given by
Eq.~\eqref{eq:power}.

At 211~keV the fluctuation lies on the resonance lines in both branches. At 99~keV the
co-passing fluctuation also lies on the resonance lines, where they cross the orbits that
reach the mode. The counter-passing response at 99~keV is broader and weaker (peak
$\widetilde F_{\rm CoM}\simeq0.7$), and the part carried by deeply passing orbits
($\lambda\lesssim0.4$) lies away from the resonance lines. This difference appears only at
the lower energy and its origin is left for future work.

\begin{figure}[!htbp]
    \centering
    \begin{subfigure}[b]{0.49\textwidth}
        \centering
        \includegraphics[width=\linewidth]{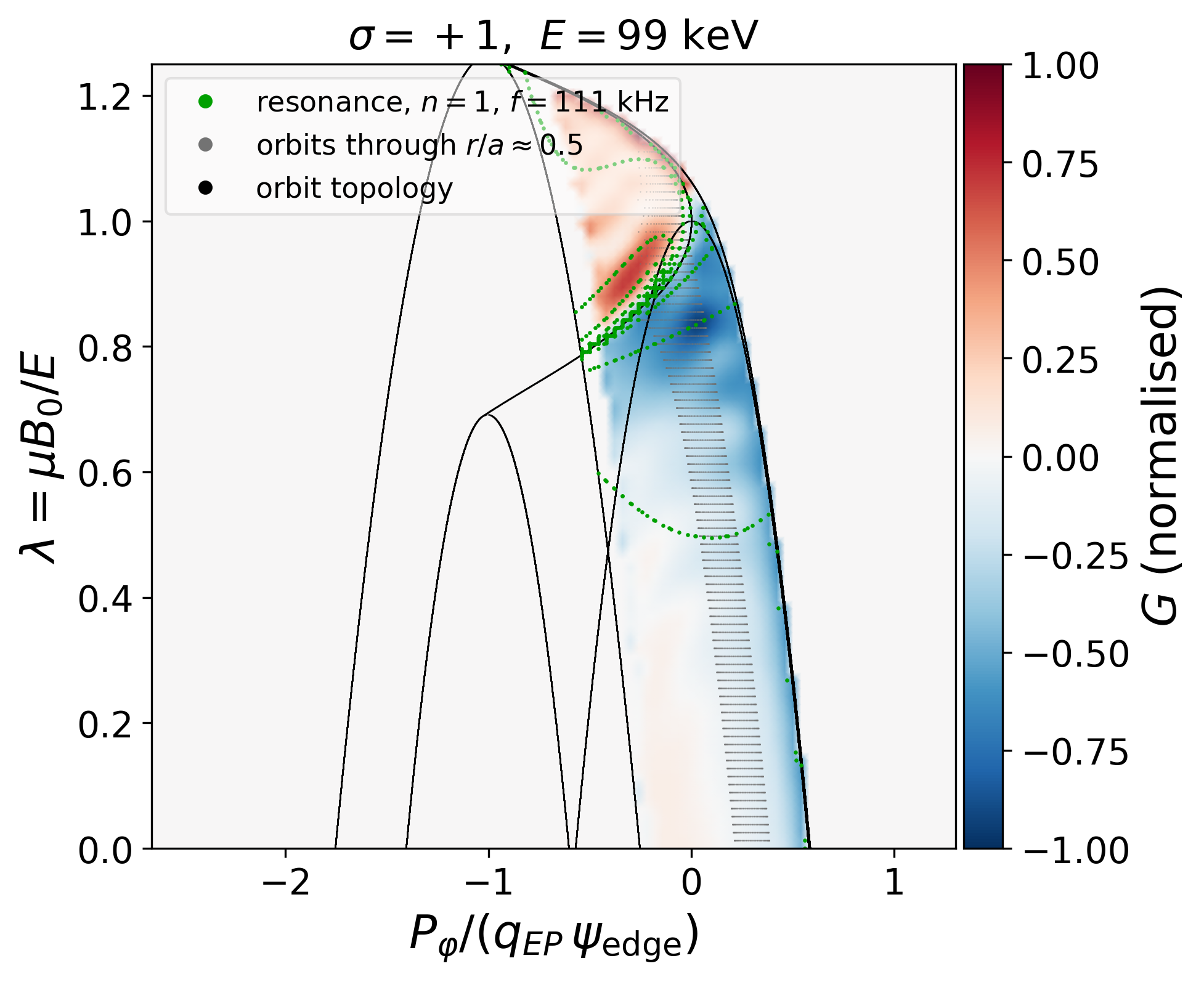}
        \caption{$G$, $\sigma=+1$}
        \label{fig:nled_G_co}
    \end{subfigure}
    \hfill
    \begin{subfigure}[b]{0.49\textwidth}
        \centering
        \includegraphics[width=\linewidth]{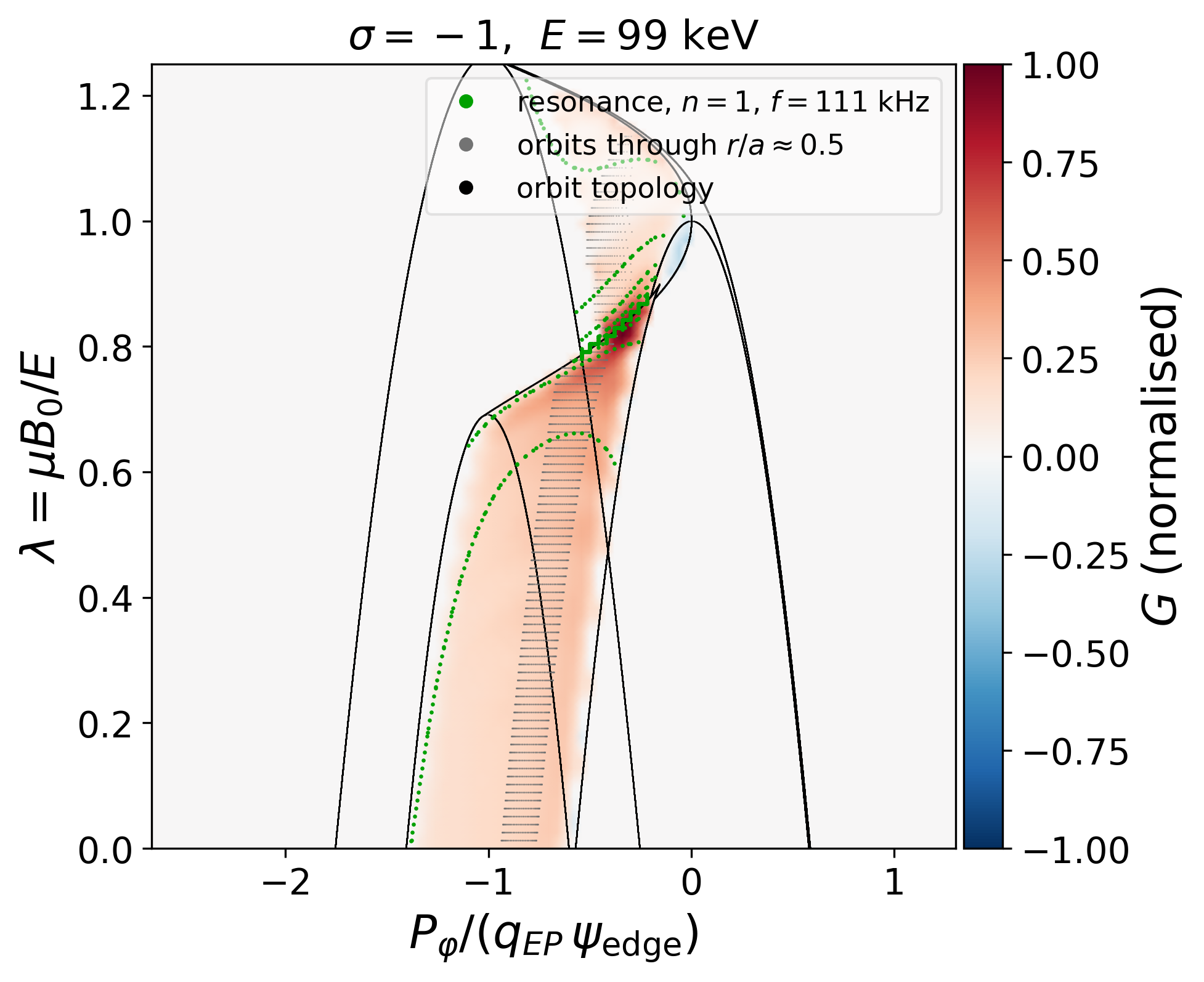}
        \caption{$G$, $\sigma=-1$}
        \label{fig:nled_G_counter}
    \end{subfigure}

    \begin{subfigure}[b]{0.49\textwidth}
        \centering
        \includegraphics[width=\linewidth]{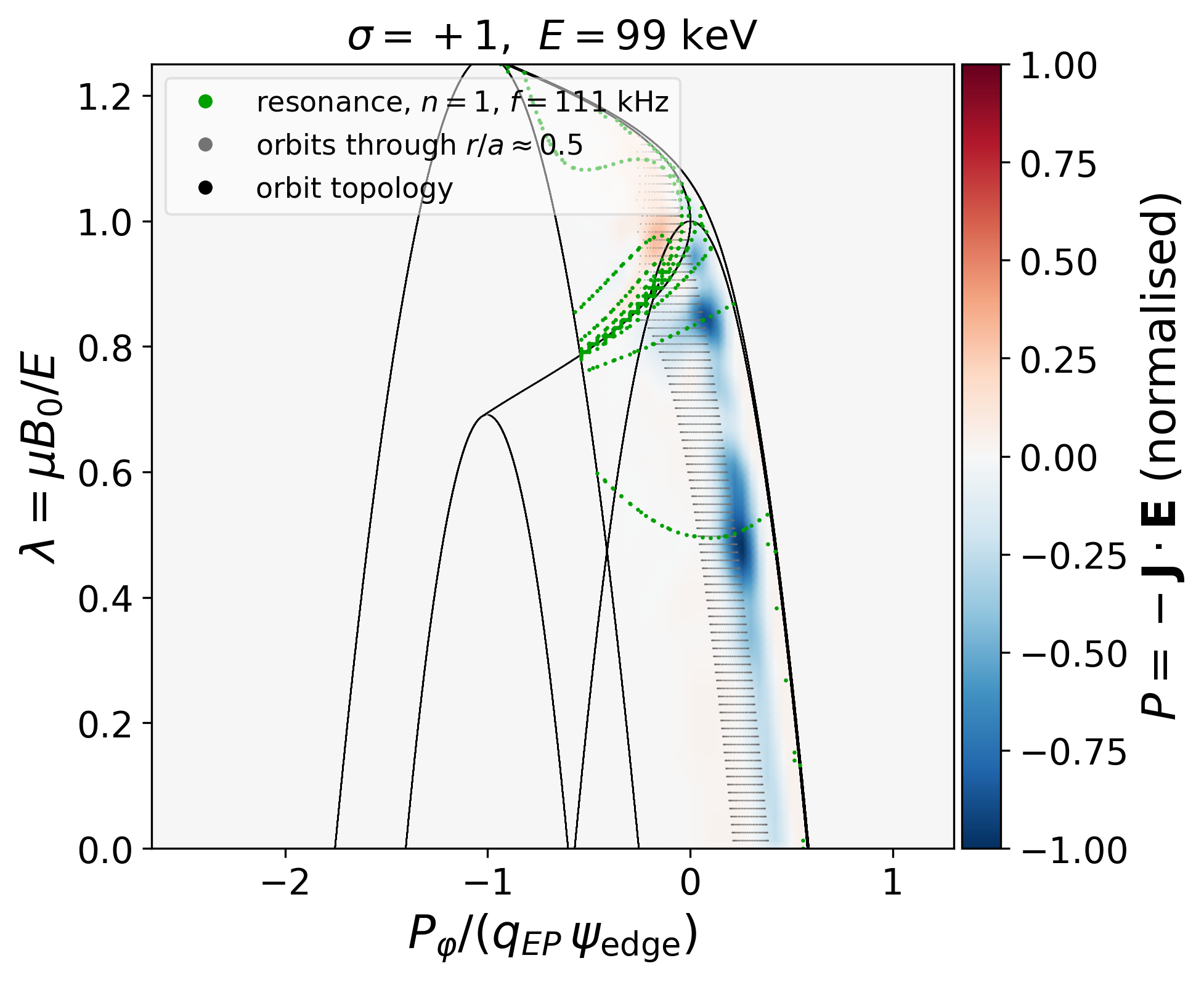}
        \caption{$P=-\mathbf J\cdot\mathbf E$, $\sigma=+1$}
        \label{fig:nled_JE_co}
    \end{subfigure}
    \hfill
    \begin{subfigure}[b]{0.49\textwidth}
        \centering
        \includegraphics[width=\linewidth]{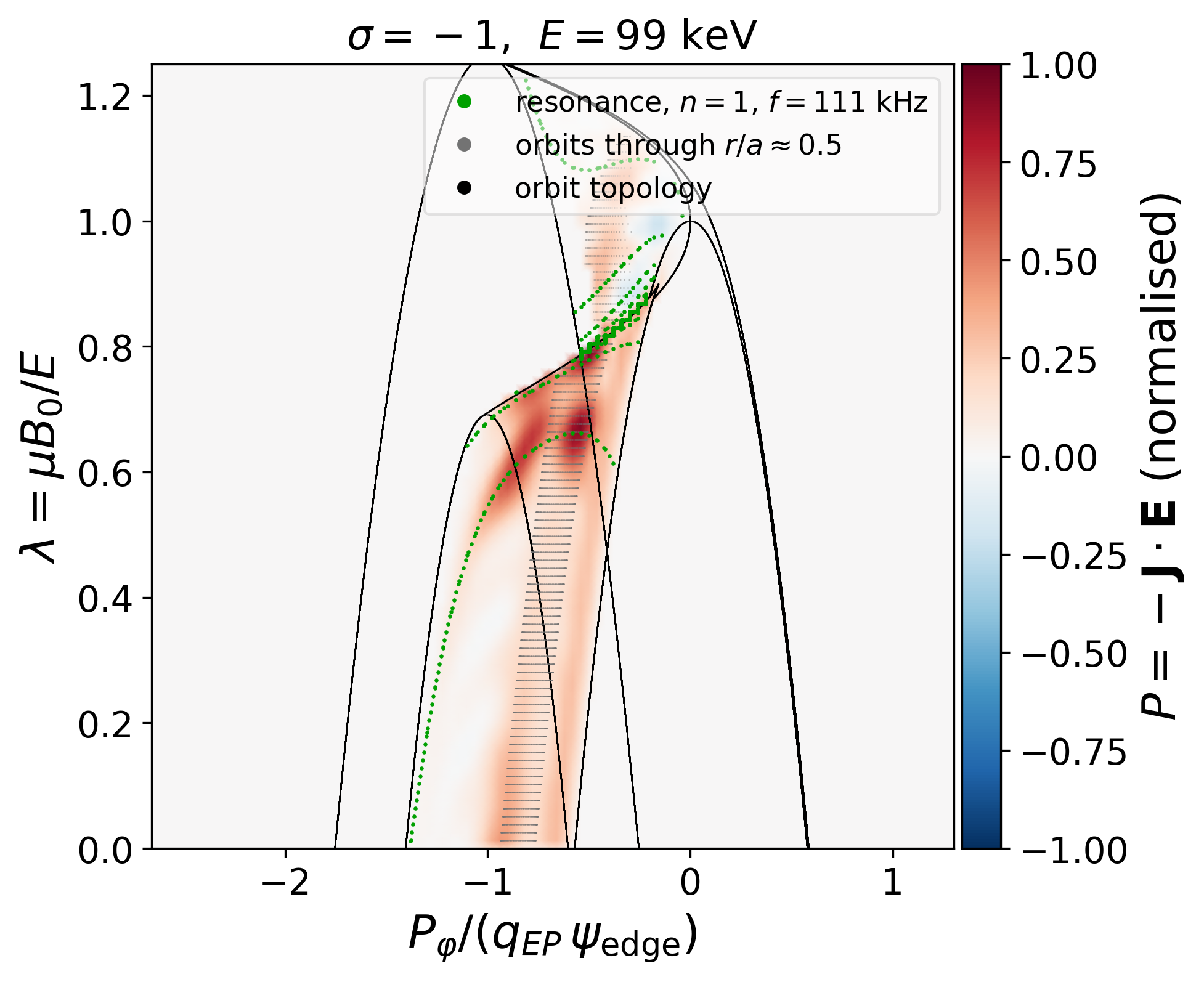}
        \caption{$P=-\mathbf J\cdot\mathbf E$, $\sigma=-1$}
        \label{fig:nled_JE_counter}
    \end{subfigure}
    \caption{EP drive $G$ (Eq.~\eqref{eq:drive_G}) (a, b) and power exchange
    $P=-\mathbf J\cdot\mathbf E$ (Eq.~\eqref{eq:power}) (c, d) at $E=99$~keV for the
    NLED-AUG case, co-passing ($\sigma=+1$) and counter-passing ($\sigma=-1$). $G$ is computed from the projected GYSELA distribution after a smoothing over one grid cell. Each panel is
    normalised to its own maximum absolute value; red: drive, blue: damping. Green dots: $n=1$ resonances
    of Eq.~\eqref{eq:res_trapped} at $f=111$~kHz; grey dots: orbits reaching
    $r/a\simeq0.5$; black dots: orbit-topology boundaries.}
    \label{fig:nled_drive}
\end{figure}

\paragraph{Drive and damping in CoM space.}
With Eq.~\eqref{eq:power}, $P>0$ means that the energetic particles give energy to the
mode (drive), and $P<0$ that they take energy from it (damping).
According to linear theory, the drive or damping from the EP distribution in CoM space is
given by~\cite{ChenZonca2016}
\begin{equation}
G = n\,\frac{\partial F}{\partial P_\varphi}
  + \omega\left(\frac{\partial F}{\partial E} - \frac{\lambda}{E}\frac{\partial F}{\partial\lambda}\right),
\label{eq:drive_G}
\end{equation}
with $G>0$ for drive and $G<0$ for damping. Fig.~\ref{fig:nled_drive} compares $G$ with
the measured exchange $P$ at $E=99$~keV. For counter-passing orbits $G>0$, and $P$ is
positive in the same region: these orbits drive the mode. For co-passing orbits $G<0$,
and $P$ is negative in the same region, these orbits damp the mode. The sign of the
exchange thus follows the sign of $G$, although the radial EP density gradient is
negative everywhere. $G$ is broad in CoM space, while $P$ is concentrated in narrower
bands, where the orbits are resonant with the mode and reach its location.
\\
The power is computed from one snapshot, at $t=8000\,\Omega_{ci}^{-1}$, after
$N_e=\gamma\,\Delta t\approx3$ e-foldings of linear growth, where $\Delta t$ is the
duration of the linear phase. A transient part from the initial condition, which does
not grow with the mode, is then about $e^{-N_e}\simeq5\%$ of the resonant exchange at the
resonance, and smaller away from it. The growth of the mode also
broadens each resonance to a width of about $\gamma/2\pi\simeq9$~kHz.

\section{Conclusions} 
\label{sec:conclusions}
We have presented linear benchmarks of energetic-particle-driven Alfv\'en eigenmodes with the
global full-$f$ gyrokinetic code GYSELA in a monotonic-shear (ITPA-TAE) and a reversed-shear
(NLED-AUG) configuration. To our knowledge, this is the first time these benchmarks have been
performed with all species (thermal ions, kinetic electrons and energetic particles) treated
full-$f$ in an electromagnetic gyrokinetic code. The agreement with the published multi-code
results is quantitative for the ITPA-TAE case and qualitative for the NLED-AUG case.

A canonical Maxwellian initialisation strongly reduces the relaxation of the fast-ion
distribution that arises in global full-$f$ models, so that the drive in the linear phase is
close to the prescribed one. In the ITPA-TAE case, GYSELA reproduces the TAE gap-mode structure
with toroidal coupling between adjacent poloidal harmonics. In the NLED-AUG case, the
instability is an $n=1$, $m=2$ RSAE localised near $q_{\min}$, with a frequency within $1\%$ of
the $m=2$ continuum accumulation point and well above the local geodesic (BAE) frequency. The
offset in frequency with respect to the reference codes is attributed mainly to the analytical
equilibrium. The density and temperature scans reproduce the instability thresholds and the
parametric trends found by the reference codes.

Beyond verification, a constants-of-motion (CoM) phase-space diagnostic was used to project
both the distribution function and the local power exchange ($P=-\mathbf J\cdot\mathbf E$) onto
invariant space, and to identify the orbit classes that drive or damp the mode. For the ITPA-TAE
case, deeply co-passing energetic particles intersect the steepest phase-space gradients and
dominate the drive. For the NLED-AUG RSAE, counter-passing orbits provide most of the drive,
with a small contribution from trapped orbits, while co-passing orbits damp the mode and absorb
about $44\%$ of the counter-passing drive. In CoM space, the sign of the exchange follows the
sign of the linear EP drive $G$, even though the radial EP density gradient is negative
everywhere. The exchange is concentrated in narrow bands, broadened by the growth rate, where
the orbits reach the mode location and are resonant with it, and the transient from the initial
condition is small at the time of the analysis. The counter-passing drive is largest near $E\simeq211$~keV, where the fluctuation lies on the $n=1$ resonance lines in both branches. At $E=99$~keV, the main counter-passing drive also lies on the resonance lines, but part of the response carried by deeply passing orbits lies away from them, and its origin is left for future work.

These results show that GYSELA can resolve which orbits drive or damp an Alfv\'enic mode in
phase space, and how much. They provide a basis for future global full-$f$ studies of the
nonlinear regime, including energetic-particle transport, Alfv\'enic mode saturation and the
role of zonal flows, for which the CoM diagnostic provides a natural framework.

\ack{%
This work is supported in part by the Ministry of
Education (MOE) AcRF Tier 1 grants RS02/23 and
RG156/23, and the National Research Foundation
Singapore (NRF) core funding ``Fusion Science for
Clean Energy'', under the framework of the Singapore
Alliance with France for Fusion Energy (SAFE).
The computational work for this article was partially
performed on resources of the National Supercomputing
Centre (NSCC), Singapore (Project 12003525).
This project was provided with computer and storage resources by GENCI at  TGCC  and  CINES, thanks to the grants 2024-A0160502224, 2025-A0180502224, and 2026-A0200502224, on the supercomputer Joliot Curie’s SKL and ROME partitions.
\\
\textit{Declaration of generative AI and AI-assisted
technologies in the manuscript preparation process:}
During the preparation of this work the authors used
ChatGPT, Gemini, and Claude in order to assist with language
polishing and used Claude to help debug the phase-space projection script.
Codex was used to adapt the manuscript to the journal template in \LaTeX.
After using these tools, the authors reviewed and edited the content as needed and take
full responsibility for the content of the published
article.
}

\data{The data that support the findings of this study are available from the corresponding author upon reasonable request.}

\appendix
\section{Fourier-based field solver in shaped tokamak geometry}
\label{app:FFT2D}
In this appendix, we introduce the Fourier-based solver for the field equations.
A spline-based solver for general tokamak geometry is already implemented in GYSELA~\cite{donnel2026adaptive}.
Nevertheless, GYSELA needs a Fourier filter on the parallel mode number to ensure numerical stability due to the CFL condition~\cite{Qu2026EMGYSELA}.
This filtering cannot be performed after the field solve as post-processing, but must be directly implemented in the linear operator by eliminating the corresponding degrees of freedom~\cite{mcmillan_rapid_2010}.
While the filtering can be done on splines coefficients~\cite{Lu2023PPCF}, it is currently incompatible with the spline library being called in GYSELA.
Therefore, we have implemented a new FFT-based solver for the quasi-neutrality and the Amp\`ere equations with in-built filtering by eliminating the associated Fourier modes in the degrees of freedom.

In general, both \eqref{eq:quasineutrality} and \eqref{eq:ampere_mixed} can be rewritten as
\begin{equation}
    -\nabla \cdot \alpha(r,\vartheta)\nabla_\perp \Phi + \beta(r,\vartheta)\Phi = \rho(r,\vartheta),
\end{equation}
where $r$ and $\vartheta$ are the generalised radius and poloidal angle, and $\alpha$ and $\beta$ are position dependent coefficients, while $\rho$ stands for charge or current on the right-hand-side.

First, we multiply both sides by $J/(R_0r)$, where $J$ is the spatial Jacobian, giving
\begin{equation}
    -\frac{J}{rR_0}\nabla \cdot \alpha(r,\vartheta)\nabla_\perp \Phi + \frac{J}{rR_0}\beta(r,\vartheta)\Phi = \mathcal{L} \Phi= \frac{J}{rR_0}\rho.
\end{equation}
Since $J\sim r$ near the axis, $J/r$ is non-singular.
The purpose of this operation is to match the 1D solver in circular geometry where the approximation $J\sim rR_0$ applies.
With the approximation that $\nabla_\perp \approx\nabla_{poloidal}$, we get
\begin{equation}
    \frac{J}{R_0r}\nabla \cdot \alpha\nabla_\perp \Phi  =\frac{1}{R_0r} \left( \partial_r Jg^{11} \alpha\partial_r + \partial_rJg^{12} \alpha \partial_\vartheta+\partial_\vartheta Jg^{12} \alpha \partial_r + \partial_\vartheta Jg^{22}  \alpha\partial_\vartheta \right) \Phi .
\end{equation}

After some algebra, we get
\begin{eqnarray}
    \frac{J}{rR_0}\nabla \cdot \alpha\nabla_\perp \Phi &=& \alpha g^{11}\frac{J}{R_0r} \left( \frac{\partial^2}{\partial r^2} + \frac{1}{r}\frac{\partial}{\partial r} \right) \Phi 
    + 2\alpha g^{12}\frac{J}{R_0r} \frac{\partial^2\Phi}{\partial r\partial\vartheta} + \alpha g^{22}\frac{J}{R_0r} \frac{\partial^2 \Phi}{\partial \theta^2}
    \nonumber\\
    &+&  \left[ \partial_r \left( \frac{J g^{11} \alpha}{rR_0}\right) + \partial_\vartheta \left( \frac{J g^{12} \alpha}{r R_0} \right)\right] \frac{\partial \Phi}{ \partial r} \nonumber\\
    &+& \left[ \frac{1}{R_0r} \partial_r\left( J g^{12} \alpha\right) + \partial_\vartheta \left( \frac{J g^{22} \alpha}{r R_0}\right)\right] \frac{\partial \Phi}{ \partial \vartheta}.
\end{eqnarray}

We use finite differences in the $r$ direction and the pseudo-spectral method in the $\vartheta$ direction.
The potential $\Phi$ is discretised as $\Phi(r_i)=\sum_{m,n}\Phi^{m,n}_{i}e^{i m \vartheta + i n \varphi}$, where $r_i$ are the grid points in $r$ and $m,n$ are the Fourier mode number is in $m$ and $n$.
We can then have the following discrete operators for the partial derivatives given by
\begin{align}
    \frac{\partial^2 \Phi}{\partial r^2} \rightarrow& \frac{\Phi_{i-1}^{m,n} - 2\Phi_{i}^{m,n} + \Phi_{i+1}^{m,n}}{\Delta r^2},\\
    \frac{\partial^2 \Phi}{\partial \vartheta^2} \rightarrow& - m^2\Phi_{i}^{m,n} \\
    \frac{\partial^2 \Phi}{\partial r \partial\vartheta} \rightarrow&  im\frac{\Phi_{i+1}^{m,n} - \Phi_{i-1}^{m,n}}{2 \Delta r}, \\
    \frac{\partial \Phi}{\partial r} \rightarrow& \frac{\Phi_{i+1}^{m,n} - \Phi_{i-1}^{m,n}}{2 \Delta r}, \\
    \frac{\partial \Phi}{\partial \vartheta} \rightarrow& im\Phi_{i}^{m,n}, \\
    \Phi \rightarrow& \Phi_{i}^{m,n},
\end{align}
while the rest are evaluated at grid points $r=r_i$.
When the geometrical axis is included, the innermost grid point is positioned at $r=\Delta r/2$.
Noting that $\Phi_{-1}^{m,n} = (-1)^m \Phi_{1}^{m,n}$ after crossing the hole, therefore
\begin{align}
        \frac{\partial^2 \Phi}{\partial r^2} + \frac{1}{ r} \frac{\partial \Phi}{\partial r}\rightarrow& \frac{- 2\Phi_{1}^{m,n} + 2 \Phi_{2}^{m,n}}{\Delta r^2},\\
        \frac{\partial \Phi}{\partial r} \rightarrow& \frac{\Phi_{2}^{m,n} - (-1)^m\Phi_{1}^{m,n}}{2 \Delta r}, \\
            \frac{\partial^2 \Phi}{\partial r \partial\vartheta} \rightarrow&  im\frac{\Phi_{2}^{m,n} - (-1)^m\Phi_{1}^{m,n}}{2 \Delta r}. 
\end{align}
For the Dirichlet boundary condition, it would simply mean $\Phi_{i}^{m,n} = 0$ if the amplitude of the mode is set to zero at radial location $r=r_i$.
The corresponding $\Phi_{i}^{m,n}$ is then dropped from computing the operator.

In this version of the FFT solver, we never explicitly construct the matrix of the discretised operator, but rather just the matrix-vector product.
To compute $\mathcal{L}\Phi$ given all the $\Phi_{i}^{m,n}$, we follow the following steps
\begin{enumerate}
    \item Apply the discrete partial differential operator as above.
    \item Transform the computed partial differentials back into real space.
    \item Multiply them by the profile and metrics.
    \item Transform back to Fourier space.
    \item Apply filter.
\end{enumerate}
The right-hand-side $J\rho/(r R_0)$ is similarly discretised by its Fourier components at radius $r_i$ given by $(J\rho/(rR_0))_i^{m,n}$.
We can then use an iterative solver such as GMRES to solve the system.

A preconditioner is needed to accelerate the convergence, which can be provided by the original GYSELA FFT solver for circular geometry.
To see this, we let, $g^{11}=1$, $g^{12}=0$, $g^{22}=1/r^2$, $J\approx rR_0$, $\alpha \approx \langle \alpha \rangle_{FS}$, which then gives
\begin{align}
    \nabla \cdot \alpha\nabla_\perp \Phi = \alpha  \left( \frac{\partial^2}{\partial r^2} + \frac{1}{r}\frac{\partial}{\partial r} \right) \Phi 
     + \alpha \frac{1}{r^2} \frac{\partial^2 \Phi}{\partial \theta^2}
    + \partial_r \alpha  \frac{\partial \Phi}{ \partial r},
\end{align}
which gives back the circular solver documented in the GYSELA literature~\cite{Grandgirard2016}.
The matrix for the circular solver is tridiagonal.
We can therefore apply the standard Thomas algorithm for a tridiagonal system.

\section{Constants-of-motion distribution of the ITPA-TAE energetic particles}
\label{app:com_distribution}
\setcounter{figure}{0}
\renewcommand{\thefigure}{B\arabic{figure}}

\begin{figure}[!htbp]
    \centering
    \includegraphics[width=\linewidth]{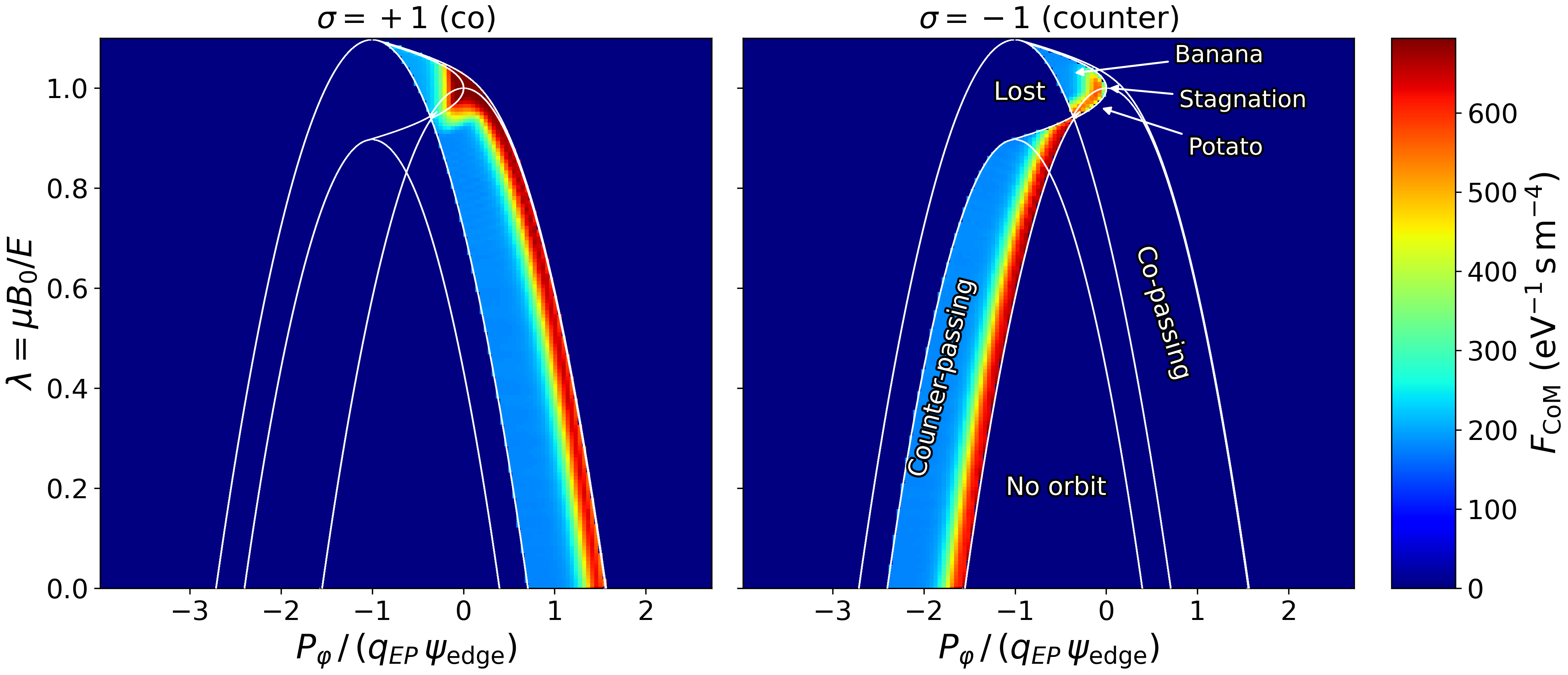}
    \caption{Constants-of-motion distribution $F_{\mathrm{CoM}}(P_\varphi,\lambda)$ of the energetic particles at $E = 407$~keV and $T_{\mathrm{EP}} = 400$~keV, for $\sigma = +1$ and $\sigma = -1$. White curves represent orbit-topology boundaries from EPCoM, with the orbit classes labelled.}
    \label{fig:FCoM_ITPA}
\end{figure}

Fig.~\ref{fig:FCoM_ITPA} shows the distribution of the energetic particles projected onto CoM space with Eq.~(\ref{eq:CoM_Projection}), at $T_{\mathrm{EP}} = 400$~keV, separately for $\sigma = +1$ and $\sigma = -1$. The white curves are the orbit-topology boundaries computed by EPCoM which separate co-passing, counter-passing, trapped and other orbit classes, and are also shown in Fig.~\ref{fig:com_fluct_reso}.

\bibliographystyle{iopart-num}
\bibliography{references}
\end{document}